\documentclass[10pt]{iopart}
\usepackage{iopams} 
\usepackage[english]{babel}

\usepackage{subfigure}
\usepackage{bm}
\usepackage{graphics}
\usepackage{graphicx}
\usepackage{iopams} 
\usepackage{float}

\usepackage[utf8x]{inputenc}
\usepackage[T1]{fontenc}
\usepackage{ae,aecompl}

\makeatletter
\renewcommand\subsubsection{\@startsection{subsubsection}{3}{\z@}%
                                     {-3.25ex\@plus -1ex \@minus -.2ex}%
                                     {1em \@plus .2em}%
                                     {\reset@font\normalsize\itshape}}
\makeatother

\usepackage[usenames,dvipsnames]{xcolor}

\usepackage{color}
\definecolor{darkgreen}{rgb}{0,0.6,0}
\definecolor{darkblue}{rgb}{0,0,0.6}
\definecolor{darkred}{rgb}{0.6,0,0}
\definecolor{darkpurple}{rgb}{0.5,0,0.5}

\usepackage[hyperindex,breaklinks,linktocpage=false]{hyperref}

\hypersetup{
bookmarksopen=true,
pdftitle=Rules of calculus in the path-integral representation of white-noise Langevin equations: the Onsager--Machlup approach,
pdfauthor=L F Cugliandolo V Lecomte, 
pdftoolbar=false, 
pdfstartview={FitH},		
pdfmenubar=true,			
pdfhighlight=/O,			
colorlinks=true,			
urlcolor=darkblue,
citecolor=darkblue,		
linkcolor=darkpurple	
}

\newenvironment{align}
 {\begin{eqnarray}}
 {\end{eqnarray}\ignorespacesafterend}

\providecommand{\paragraphitalics}[1]{%
\paragraph{\textnormal{{\it \thesubsubsection.#1}.~}}%
\addcontentsline{toc}{subsubsection}{\thesubsubsection.#1}%
}

\newcommand{\tf}{t_{\text{f}}}
\newcommand{\st}{{\text{\tiny{S}}}}
\newcommand{\xs}{\bar x^{\st}}

\newcommand{\dt}{\text{d}_t}

\newcommand{\ee}{\text{e}}

\newcommand{\GG}{\text{G}}

\newcommand{\J}{\mathcal J}
\newcommand{\PP}{\mathbb P}
\newcommand{\Prob}{\text{Prob}}

\renewcommand{\phi}{\varphi}

\providecommand{\tfrac}[2]{\textnormal{$\frac{#1}{#2}$}}
\providecommand{\eqref}[1]{\eref{#1}}
\providecommand{\text}[1]{{\rm{#1}}}

\begin{document}


\title{%
Rules of calculus in the path integral representation of white noise Langevin equations:\\ the Onsager--Machlup approach
}


\author{Leticia F.~Cugliandolo$^{1}$ and Vivien~Lecomte$^{2,3}$}

\address{$^1$
Sorbonne Universités, Université Pierre et Marie Curie - Paris VI, Laboratoire de Physique Théorique et
Hautes Énergies (LPTHE), 4 Place Jussieu, 75252 Paris Cedex 05, France
}

\address{$^2$
LIPhy, Université Grenoble Alpes \& CNRS,  F-38000 Grenoble, France
}
\address{$^3$
Laboratoire Probabilités  et Modèles Aléatoires (LPMA), CNRS UMR 7599, Université Paris Diderot, Paris Cité Sorbonne, Bâtiment Sophie Germain, Avenue de France, F-75013 Paris, France}

\ead{\\leticia@lpthe.jussieu.fr\\ vivien.lecomte@univ-grenoble-alpes.fr}

\vspace{10pt}


\begin{abstract}
The definition and manipulation of Langevin equations with multiplicative white noise require special care (one has to specify the time discretisation and a 
stochastic chain rule has to be used to perform changes of variables). 
While discretisation-scheme transformations and  non-linear changes of variable can be safely performed
on the Langevin equation, 
these same transformations lead to inconsistencies in its
path-integral representation. We identify their origin and we 
show how to extend the well-known   Itō prescription 
($dB^2 = dt$) in a way that defines a modified stochastic calculus to be used inside the path-integral representation of the 
process, in its Onsager-Machlup form.
\end{abstract}

\vspace{2pc}


\bigskip


\newpage
\begin{small}
	\tableofcontents
\end{small}

\newpage
~\bigskip
\section{Introduction}
\label{sec:introduction}

Physical phenomena are often non-deterministic, presenting a stochastic behaviour induced by the action of a large number of 
constituents or by more intrinsic sources of noise~\cite{Stratonovich92,gardiner_handbook_1994,kampen_stochastic_2007,oksendal_stochastic_2013}.
A paradigmatic example is the one of Brownian motion, the study of which is at the source of stochastic calculus.
From a modelisation viewpoint, the evolution of such systems can be described by a Langevin-type equation or by the path probability of its trajectories.
An important aspect of these descriptions is that the trajectories are not differentiable in general. This peculiarity
implies that the definition of the evolution equation  requires 
special care, namely, it demands the specification of a non-ambiguous time-discretisation
scheme and, moreover, it induces a modification of the rules of calculus~\cite{Stratonovich92,gardiner_handbook_1994,kampen_stochastic_2007,oksendal_stochastic_2013}.

The important role of the time discretisation in the Langevin equation is now clearly elucidated~\cite{kampen_ito_1981}
and many results have been obtained for the construction of an associated path-integral formalism, whose functional action and Jacobian correctly take 
into account the choice of discretisation
\cite{lau_state-dependent_2007,aron_dynamical_2014_arxiv1,aron_dynamical_2016,Aron_etal_2014,itami_universal_2017,arenas_functional_2010,moreno_langevin_2015,barci_path_2016}.

An important point in the manipulation of Langevin equations is that the usual differential-calculus chain rule for changes of variables,
$\dt\big[u(x(t))\big]=u'(x(t))\, \dt x(t)$,
has to be modified.
It is replaced by the Itō formula (or `stochastic chain rule'), which is itself the consequence of the Itō substitution rule $dB^2=dt$
for an infinitesimal increment $dB=B_{t+dt}-B_t$ of a Brownian motion $B_t$ of unit variance.
Although such manipulations are well understood at the Langevin equation level,  
the situation is less clear for the transformation of fields performed \emph{inside} the action functional  corresponding to the Langevin equation.
It is known, for instance, that the use of the stochastic chain rule in the action can yield unsolved inconsistencies, both in statistical field theory~\cite{Tirapegui82,aron_dynamical_2014_arxiv1} and in quantum field theory~\cite{gervais_point_1976,Sa77,LaRoTi79,Tirapegui82,AlDa90,ApOr96}.

In this article, we elucidate the source of this inconsistency, focusing on the case of the Onsager-Machlup action functional corresponding to a Langevin equation for one degree of freedom, with multiplicative white noise.
We find that the sole Itō substitution rule $dB^2=dt$ proves to be insufficient  
to correctly perform non-linear changes of variables in the action. We identify the required generalised substitution rules
and we determine that their use should be performed with extreme care, since they take different forms
when applied inside the exponential of the time-discrete action, or in the prefactor of its Gaussian weight factor.

In continuous time, we show that, in general, the use of the usual stochastic chain rule inside the action 
yields wrong results --~and this even for a Stratonovich-discretised additive-noise Langevin equation.
We determine a modified stochastic chain rule that allows one to manipulate the action directly, even in continuous time.

The organisation of the article is the following.
In Sec.~\ref{sec:def-models}, we review the non-ambiguous construction of the  Langevin equation, 
providing three detailed examples which illustrate the role of the Itō substitution rule.
In Sec.~\ref{sec:stoch-calc-path}, we recall inconsistencies that appear when one manipulates the action incorrectly, and we 
determine the valid substitution rules.
We synthesise our results in Sec.~\ref{sec:outlook}.
Appendices gather part of the technical details.


\section{Langevin equation and stochastic calculus}
\label{sec:def-models}

In this Section, we briefly review the definition of multiplicative Langevin equations. For completeness, we first describe the standard construction of an 
unambiguous stochastic evolution equation through time discretisation, 
and we then provide three examples illustrating how differential calculus is generalised for stochastic variables, following this construction.

\subsection{Discretisation convention of Langevin equations}
\label{sec:discr-conv}

Consider a time-dependent 
variable $x(t)$ which verifies a Langevin equation with a force $f(x)$ and a multiplicative noise $g(x)\eta$,
\begin{align}
  \label{eq:langevin}
  \dt x(t) = f(x(t))+g(x(t))\,\eta(t)
  \; .
\end{align}%
The function $g(x)$, that depends in general on the value of the variable, 
describes the amplitude of the stochastic term of this equation. 
The noise $\eta(t)$ is a centred Gaussian white noise of 2-point correlator equal to
$
  \langle\eta(t)\eta(t')\rangle=2D\delta(t'-t)
$,
where $D$ plays the role of temperature.
It is well known that the Langevin equation in its continuous-time writing~(\ref{eq:langevin}) is ambiguous: one needs to specify a `discretisation scheme' in order to give it a meaning (see \cite{kampen_ito_1981,kampen_stochastic_2007} for reviews).

Such a scheme is defined in discrete time, in the zero time step limit.
We denote by $x_t$ the time-discrete variable, with now $t\in dt\,\mathbb N$.
The central feature of the definition of the Langevin equation is the following.
Upon the time step $t \curvearrowright t+dt$, the right-hand-side (r.h.s.) of~(\ref{eq:langevin}) is evaluated at a 
value of $x=\bar x_t$ \emph{chosen as a weighted average} between $x_t$ and $x_{t+dt}$ as
\begin{align}
  \label{eq:langevindiscalpha}
  \frac{x_{t+dt}-x_t}{dt} &= f(\bar x_t)+g(\bar x_t)\eta_t
  \qquad\qquad
  \langle\eta_t\eta_{t'}\rangle= \frac{2D}{dt}\,\delta_{tt'}
\end{align}
where the $\alpha$-discretised evaluation point is
\begin{align}
  \label{eq:langevindiscalpha_2}
  \bar x_t &= \alpha x_{t+dt}+(1-\alpha)x_t=x_t+\alpha(x_{t+dt}-x_t)
  \; . 
\end{align}
In the time-discrete evolution~(\ref{eq:langevindiscalpha}), the noise $\eta_t$ is a centred Gaussian random variable (independent from those at other times, and thus independent of $x_t$). Its explicit distribution reads
\begin{align}
  \label{eq:dist}
  \forall t,\quad P_\text{noise}(\eta_t) = \sqrt{\frac{dt}{4\pi D}}\: \ee^{-\frac 12 \frac{dt}{2D} {\eta_t^2}}
  \; . 
\end{align}
Its form implies that the stochastic term~$g(\bar x_t)\eta_t$ in (\ref{eq:langevindiscalpha}) is 
typically of order $dt^{-1/2}$,  that is much larger than $f(\bar x_t)$, which is  of order $dt^0$.
This difference is at the core of the ambiguity of the equation~(\ref{eq:langevin}): 
as $dt\to 0$, the deterministic contribution $f(\bar x_t)$ to~(\ref{eq:langevindiscalpha}) is independent of the choice of $\alpha$-discretisation; 
however, different values of $\alpha$ lead to different behaviours of the stochastic term~$g(\bar x_t)\eta_t$.
Indeed, making  the discretisation explicit with a superscript we see, by Taylor expansion, that 
\begin{align}
  \fl\qquad
  \Big[
  g\big(\bar x_t^{(\bar\alpha)}\big)-g\big(\bar x_t^{(\alpha)}\big)
  \Big]
  \eta_t
  \nonumber
&=
  \Big[
  g\big(\bar x_t^{(\alpha)}+(\bar\alpha-\alpha)(x_{t+dt}-x_t)\big)-g\big(\bar x_t^{(\bar\alpha)}\big)
  \Big]
  \eta_t
\\
&=(\bar\alpha-\alpha)(x_{t+dt}-x_t)g'(\bar x_t^{(\alpha)})\eta_t+O(dt^{-\frac 12})
\label{eq:noise-difference_alphaalphabar}
\end{align}
is typically of order $dt^0$. This shows that, in general, $g(\bar x_t^{(\alpha)})\eta_t$ and $g(\bar x_t^{(\bar \alpha)})\eta_t$ are not equivalent in~(\ref{eq:langevindiscalpha}) when $\alpha\neq\bar\alpha$.

Standard discretisation choices are $\alpha=\frac 12$ (known as `mid-point' or Stratonovich convention) and $\alpha=0$ (Itō convention).
The Stratonovich choice is invariant by time reversal but, as other choices of $0<\alpha\leq 1$, yields an implicit equation~(\ref{eq:langevindiscalpha}) for $x_{t+dt}$ at each time step.
The Itō convention, yielding independent increments for $x(t)$, is often chosen in mathematics, where the construction of the corresponding ``stochastic calculus''~\cite{oksendal_stochastic_2013} is done by defining a stochastic integral for the integral equation corresponding to~(\ref{eq:langevin}).

In general, we will denote the $\alpha$-discretised Langevin equation~(\ref{eq:langevin}) as
\begin{align}
  \label{eq:langevin_alpha}
  \dt x \stackrel\alpha= f(x)+g(x)\,\eta
  \; . 
\end{align}

\subsection{Three examples}
\label{sec:three-examples}

In this Section, we review three archetypal situations illustrating the role played by the choice of $\alpha$-discretisation. We explain the computations in  detail, so as to start off on the right footing for understanding the origin of the apparent contradictions discussed in Sec.~\ref{sec:stoch-calc-path}.

\subsubsection{The stochastic chain rule (or Itō formula)}
\label{sec:chain-rule}

A first consequence of the presence of a term of order $dt^{-1/2}$ in the discrete-time Langevin equation~(\ref{eq:langevindiscalpha}) is that the usual formulæ of differential calculus  have to be altered.
For instance, the chain rule describing the time derivative of a function of $x(t)$ is modified as~\cite{kampen_stochastic_2007,gardiner_handbook_1994}
\begin{align}
  \label{eq:chainrule_example}
  \dt\! \big[u(x)\big] 
  \stackrel\alpha= 
  u'(x)\dt  x + (1-2\alpha)D\,g(x)^2 u''(x)
\;,
\end{align}
where $x=x(t)$ verifies the Langevin equation~(\ref{eq:langevin_alpha}).
It is only for the Stratonovich discretisation that one recovers the chain rule of differentiable functions.
For $\alpha =0$, the relation~(\ref{eq:chainrule_example}) is known as the ``Itō formula''.

The stochastic chain rule~(\ref{eq:chainrule_example}) is understood as follows. Coming back to the discrete-time definition of $\dt [u(x)]$, one performs a Taylor expansion in powers of 
$
\Delta x \equiv x_{t+dt}-x_{dt}
$,
keeping in mind that, as seen from~(\ref{eq:langevindiscalpha}), $\Delta x$ is of order $dt^{1/2}$; this yields
\begin{align}
\label{eq:discretetichainrule}
  &\frac{u(x_{t+dt})-u(x_t)}{dt}
  =
    \frac{u\big(\bar x_{t}+(1-\alpha)\Delta x\big)-u\big(\bar x_{t}-\alpha\Delta x\big)}{dt}
\nonumber\\
  &
  \qquad \qquad \qquad
  =
  \frac{\Delta x}{dt} u'(\bar x_t)
    +
    \frac 12 (1-2\alpha) \frac{\Delta x^2}{dt} u''(\bar x_t)
    + O(dt^{\frac 12})
    \; . 
\label{eq:chainrule_Dx2dt}
\end{align}
For a differentiable function $x(t)$, the term $\propto \Delta x^2/dt$ would be negligible in the $dt\to 0$ limit but this is not the case for a stochastic $x(t)$.
The next step is to understand the continuous-time limit $dt\to 0$ of~(\ref{eq:chainrule_Dx2dt}): the so-called ``Itō prescription'' amounts to replacing $\Delta x^2/dt$ in this expression by its quadratic variation
\begin{align}
  \label{eq:itopresc}
  \frac{\Delta x^2}{dt} \:\mapsto\: 2 D g(\bar x_t)^2 
  \qquad 
  \text{as~~} dt\to 0
\end{align}
(which is not equal to the expectation value of ${\Delta x^2}/{dt}$, as occasionally read in the literature, since $g(\bar x_t)$ depends on the value of $\bar x_t$ without averaging).
Note that in Eq.~\eqref{eq:chainrule_Dx2dt}, one could as well replace ${\Delta x^2}/{dt}$ by the Itō{-discretised} $ 2 D g( x_t)^2$ (or any other discretisation point) 
instead of the \mbox{$\alpha$-discretised} one  in~(\ref{eq:itopresc}) 
since this would only add terms  of order $dt^{1/2}$ to~\eqref{eq:chainrule_Dx2dt} --~hence the name ``Itō prescription''.
In this article, we will rather use the name ``substitution rule'' for two reasons: one is that we work in a generic $\alpha$-discretisation scheme; another one is that we will introduce generalisations of~(\ref{eq:itopresc}) at a later stage.

We emphasise that the substitution rule~(\ref{eq:itopresc}) has to be used with care, as will be illustrated many times in this article.
The validity of its use relies on the precise definition of the chain rule~\eqref{eq:chainrule_example}: this identity has to be 
understood in an ``$L^2$-norm'' sense, \emph{i.e.}~it 
corresponds to having $\langle[\int_0^{\tf}dt\,\{\text{l.h.s.}-\text{r.h.s.}\}]^2\rangle=0$~$(\forall \tf)$ 
and not to having a point-wise equality. The precise formulation and the demonstration of~(\ref{eq:chainrule_example}) and~(\ref{eq:itopresc})
are given in Sec.~\ref{sec:just-it=o} of App.~\ref{sec:just-gener-it=o}, along the lines of Øksendal's 
reference textbook~\cite{oksendal_stochastic_2013}.

As this sort of issues is often overlooked in the theoretical physics literature, we now explain why an argument that is regularly proposed to justify~(\ref{eq:itopresc}) is in fact invalid.
One could argue that the distribution of $\Delta x^2$ in~(\ref{eq:chainrule_Dx2dt}) is sharply peaked around its most probable value $2Dg(\bar x_t)^2dt$, because its variance $\langle \Delta x^4\rangle-\langle \Delta x^2\rangle^2$ is of order~$dt^2$ as read from~(\ref{eq:langevindiscalpha}) and~(\ref{eq:dist}); this would allow one to replace $\Delta x^2/dt$ by $2Dg(\bar x_t)^2$ as $dt\to 0$ in~(\ref{eq:chainrule_Dx2dt}), hence justifying~(\ref{eq:itopresc}).
However, this argument is incorrect because the variance of $\Delta x^2/dt$  is of the same order~$dt^0$ as some other terms in the time-discrete Langevin equation~(\ref{eq:chainrule_Dx2dt}).
To understand this point in detail, it is convenient to rephrase the argument as follows.
First, one notes that according to~(\ref{eq:langevindiscalpha}) and~(\ref{eq:dist}), the quantity ${\Delta x}/{dt}$ is dominated by its most singular contribution $g( x_t)\eta_t$ in the $dt\to 0$ limit
\begin{align}
\frac{\Delta x}{dt}=g(x_t)\eta_t+O(dt^0)
\; . 
\end{align}
In this expression, we have chosen to evaluate $g(x)$ at 
$x=x_t$ instead of $\bar x_t$, the difference being gathered with other terms of order $O(dt^0)$ (see~\eqref{eq:noise-difference_alphaalphabar} for a proof).
This allows one to use the fact that $\eta_t$ is independent of $x_t$ in order to compute the variance of $\Delta x^2/dt$  by Gaussian integration over $\eta_t$ as 
\begin{align}
\fl
\qquad
  \Big \langle \Big[ \frac{\Delta x^2}{dt}\Big]^2\Big\rangle
-
  \Big \langle \frac{\Delta x^2}{dt}\Big\rangle^2 
  \nonumber
  &
=
\Big\{
  \Big \langle \big[g(x_t)\eta_t \big]^4\Big\rangle 
-
  \Big \langle \big[g(x_t)\eta_t \big]^2\Big\rangle^2 
\Big\}\, dt^2 
+ O(dt)
\\
&
= 4 D^2 \Big( 3\langle g(x_t)^4 \rangle - \langle g(x_t)^2 \rangle^2\Big)  + O(dt)
\end{align}
%
%
and one observes that it does not vanish as $dt\to 0$ (even for a constant noise amplitude~$g(x)=g$).
The variance of $\Delta x^2/dt$ is thus of the same order $dt^0$ as other terms in 
Eq.~(\ref{eq:chainrule_Dx2dt})
; this means that the properties of the distribution of  $\Delta x^2/dt$  cannot be invoked to justify the substitution rule~(\ref{eq:itopresc}).
This rule has to be understood in an~$L^2$ sense that we explain in App.~\ref{sec:just-it=o}.
As will prove to be essential, it means that the chain rule~(\ref{eq:chainrule_example}) is not true ``point-wise'' but only in a weaker sense -- which has to be taken care of meticulously in the path integral action, as we discuss thoroughly in Sec.~\ref{sec:non-line-transf}.

Finally, we note that the substitution rule~(\ref{eq:itopresc}) is equivalently written as follows%
\footnote{%
Another writing is $dB_t^2\mapsto dt$ for a Brownian motion $B_t$ of unit variance -- the relation with our discrete white noise being 
$\eta_t\,dt=(2D)^{1/2}(B_{t+dt}-B_t)$.
} 
\begin{align}
  \label{eq:itopresc_eta}
  \eta_t^2{dt}\mapsto 2 D 
\end{align}
for the discrete time white noise $\eta_t$.

\subsubsection{Changing discretisation while keeping the same evolution}
\label{sec:chang-discr-while}

Since the solution $x(t)$ of the Langevin equation~(\ref{eq:langevin}) depends crucially on the choice of $\alpha$-discretisation, although
this choice seems to be arbitrary, one can wonder whether~
$x(t)$ can also be described as the solution of another Langevin equation, with a different $\bar\alpha$-discretisation and a modified force.
To answer this question, one comes back to the discrete-time evolution~(\ref{eq:langevindiscalpha})-(\ref{eq:langevindiscalpha_2})
\begin{align}
  \label{eq:langevindiscalpha_explicit}
  \frac{x_{t+dt}-x_t}{dt} &= f\big(\bar x^{(\alpha)}_t\big)+g\big(\bar x^{(\alpha)}_t\big)\eta_t
\end{align}
where we wrote explicitly the discretisation convention in superscript.
Then, writing
\begin{align}
  \label{eq:langevindiscalpha_2_changealpha}
  \bar x^{(\alpha)}_t 
  &= 
    \bar x_t^{(\bar\alpha)}+(\alpha-\bar\alpha)(x_{t+dt}-x_t)
\end{align}
and expanding in powers of $\Delta x=x_{t+dt}-x_t=O(dt^{1/2})$
one obtains
\begin{align}
  \label{eq:langevindiscalpha_change}
\fl\qquad
  \frac{x_{t+dt}-x_t}{dt} 
  &= 
    f\big(\bar x^{(\bar\alpha)}_t\big)+g\big(\bar x^{(\bar\alpha)}_t\big)\eta_t
    +(\alpha-\bar\alpha)g'\big(\bar x^{(\bar\alpha)}_t\big)\Delta x\,\eta_t
    +O(dt^{\frac 12})
\nonumber\\
  &= 
    f\big(\bar x^{(\bar\alpha)}_t\big)+g\big(\bar x^{(\bar\alpha)}_t\big)\eta_t
    +(\alpha-\bar\alpha)g\big(\bar x^{(\bar\alpha)}_t\big)g'\big(\bar x^{(\bar\alpha)}_t\big)\,\eta_t^2dt
    +O(dt^{\frac 12})
\end{align}
where we used~(\ref{eq:langevindiscalpha}) for the last line.

Finally, using the substitution rule~(\ref{eq:itopresc_eta}) and sending $dt$ to zero, one finds that the process $x(t)$, solution of the Langevin equation~(\ref{eq:langevin_alpha}) in the $\alpha$-discretisation, is also verifying another Langevin equation
\begin{align}
  \label{eq:langevin_alpha-to-alphabar}
  \dt x 
  &
    \stackrel{\bar\alpha}= 
    f_{\alpha\to\bar\alpha}(x)+g(x)\,\eta
\end{align}
\begin{align}
  \label{eq:langevin_alpha-to-alphabar_force}
  f_{\alpha\to\bar\alpha}(x)
  &=
    f(x)+ 2(\alpha-\bar\alpha) D\,g(x)g'(x)
\end{align}
which is understood in $\bar\alpha$-discretisation and presents a modified force $f_{\alpha\to\bar\alpha}(x)$.
One checks directly that the Fokker-Planck equations corresponding to the two Langevin equations~(\ref{eq:langevin}) and~(\ref{eq:langevin_alpha-to-alphabar})-(\ref{eq:langevin_alpha-to-alphabar_force}) are identical, illustrating 
the equivalence of the two corresponding processes (see for instance~\cite{kampen_ito_1981} for the special case $\alpha=0$ and $\bar\alpha=1/2$).
However, we emphasise that, since we used the substitution rule~(\ref{eq:itopresc_eta}), we have to keep in mind that the equivalence between~(\ref{eq:langevin_alpha}) and~(\ref{eq:langevin_alpha-to-alphabar})-(\ref{eq:langevin_alpha-to-alphabar_force}) is not true pointwise and this can be the source of unexpected problems, as discussed in Sec.~\ref{sec:from-discr-anoth}.

\subsubsection{Infinitesimal propagator for a path integral formulation}
\label{sec:infin-prop-path}

The trajectory probability of stochastic processes described by a Langevin equation has been the focus of many studies in statistical mechanics, either from the Onsager--Machlup approach
~\cite{onsager_fluctuations_1953,machlup_fluctuations_1953II} or from the Martin--Siggia--Rose--Janssen--De\,Dominicis (MSRJD) one~\cite{Martin1973,DeDominicis1976,Janssen1976,BaJaWa76,Janssen1979,DeDominicis1978}.
The idea in the Onsager--Machlup approach (to which we restrict our present analysis) is to write the probability of a trajectory $[x(t)]_{0\leq t\leq\tf}$ as
\begin{align}
  \label{eq:trajprobcont}
  \Prob\big[x(t)\big]
  =
  \J[x(t)]
  \,
\ee^{-S[x(t)]}
\; , 
\end{align}
where %
$S[x(t)]$ is the ``action'',
which takes a Lagrangian form $S[x(t)]=\int_0^{\tf}\!dt\:\mathcal L(x,\dt x)$,
and $\J[x(t)]$ is a ``normalisation prefactor''\footnote{%
The prefactor $\J[x(t)]$ can be included in the measure $\mathcal Dx$ on trajectories, but is not exponentiated in the action in general because it does not take a Lagrangian form.
}%
.
As can be expected from the discussion at the beginning of subsec.~\ref{sec:discr-conv}, the form of the action and of the normalisation prefactor will depend not only on the  $\alpha$-discretization of the underlying Langevin equation, but also on the discretisation convention which is used to write them.
The average of a functional $\mathcal F$ of the trajectory can then be written in a path integral form as
\begin{align}
  \big\langle \mathcal F\big[x(t)\big]\big\rangle
  =
  \int \mathcal Dx\,\mathcal F\big[x(t)\big]\,\J[x(t)]  \,\Prob\big[x(t)\big]
  P_{\text{i}}\big(x(0))
  \; . 
\end{align}
The path integral is understood in the Feynman sense~\cite{feynman_space-time_1948}: a sum over possible trajectories which start from an initial condition sampled by a distribution~$P_{\text{i}}(x)$. It is best depicted in a time-discrete setup in the limit of zero time step, where one integrates over the set of possible values $x_{t}$ of the trajectory at discrete times $t\in dt\,\mathbb N$ separated by a time step $dt$, yielding
\begin{align}
  \prod_{t=0}^{\tf/dt-1}
  \!\!\!
  \big\{
  dx_t\,
  \PP(x_{t+dt},t+dt|x_{t},t)
  \big\}
\quad
&  \stackrel{dt\to 0}\longrightarrow
  \quad
  \mathcal Dx
  \,
  \J[x(t)]
  \,
\ee^{-S[x(t)]}
\label{eq:Feynman-path-int}
\end{align}
where $\PP(x_{t+dt},t+dt|x_{t},t)$ is a conditional probability (or ``infinitesimal propagator'').

In this subsection, we focus our attention on the infinitesimal propagator between two successive time steps, that for simplicity we take at the first time step. Our goal is to compute
$
\mathbb P(x_{dt}|x_0)
\equiv
\PP(x_{dt},dt|x_{0},0)
$ and to understand how the full action and normalisation prefactor are reconstituted through~\eqref{eq:Feynman-path-int}.
We note that the correct form of this propagator, taking into account the $\alpha$-discretisation 
is well-known~\cite{lau_state-dependent_2007,Aron_etal_2014,aron_dynamical_2016,itami_universal_2017}. 
Still, we derive it again by taking a pedestrian approach that illustrates the role played by the 
substitution rules~(\ref{eq:itopresc}) or~(\ref{eq:itopresc_eta}) --~a role that proves 
essential to understand in order to later find the correct rules of stochastic calculus in the action.

\paragraphitalics{a First time step: changing from the distribution of $\eta_0$ to that of $x_{dt}$}
\label{ssec:first-time-step}

Let us fix the initial condition $x_0$ and determine the distribution of $x_{dt}$ obtained from the discrete Langevin equation~\eqref{eq:langevindiscalpha}. This equation is an implicit equation on $x_{dt}$, the solution of which takes the form
\begin{align}
  \label{eq:x1X1}
  x_{dt}=X_1(x_0,\eta_0)
  \; . 
\end{align}
Therefore, the distribution of $x_{dt}$ reads
\begin{align}
  \label{eq:Px1eta0}
  \mathbb P(x_{dt}|x_0) = 
  \int d\eta_0\: 
  \delta(x_{dt}-X_1(x_0,\eta_0)) \: 
  P_\text{noise}(\eta_0) \ ,
\end{align}
with the noise distribution given in Eq.~\eqref{eq:dist}.
In order to integrate over $\eta_0$, we would like to read the Dirac as a $\delta$ on the variable $\eta_0$.
Cancelling the argument of the Dirac distribution in~\eqref{eq:Px1eta0} defines a function $H_0(x_0,x_{dt})$ such that
\begin{equation}
 \big(x_{dt}-X_1(x_0,\eta_0)\big)\Big|_{\eta_0=H_0(x_0,x_{dt})}=0
 \; . 
\label{eq:defH0}
\end{equation}
Then, the relation~\eqref{eq:Px1eta0} yields
\begin{align}
  \label{eq:Px1}
  \mathbb P(x_{dt}|x_0) 
  &= \int d\eta_0\: 
    \frac
    {\delta(\eta_0-H_0(x_0,x_{dt}))}
    {\Big|\partial_{\eta_0}\big(x_{dt}-X_1(x_0,\eta_0)\big)\Big|_{ \rlap{$\mathsurround=0pt\scriptstyle{ \eta_0=H_0(x_0,x_{dt})   }$} }}
    P_\text{noise}(\eta_0)     
\nonumber\\
  &=  
    \frac
    {P_\text{noise}\big(H_0(x_0,x_{dt})\big)}
    {\Big|\partial_{\eta_0}X_1\big(x_0,H_0(x_0,x_{dt})\big)\Big|}
    \;  .
\label{eq:Px1x0Pnoise}
\end{align}
Note that this relation can be derived by performing a change of variables 
in the probability distribution $P_{\text{noise}}$ of $\eta_0$, to obtain the distribution of $x_{dt}$ seen as a function of $\eta_0$ through~\eqref{eq:x1X1}.
(Two ways of evaluating the denominator are recalled in App.~\ref{sec:anoth-comp-jacob-IS}~\cite{itami_universal_2017} and 
App.~\ref{sec:anoth-comp-jacob-Arnold}~\cite{aron_dynamical_2016}; we follow here
a different route that is better adapted for our purposes.)

\paragraphitalics{b Expansions in the limit $dt\to 0$} 
The discrete Langevin equation~\eqref{eq:langevindiscalpha} relating (at $t=0$) $x_0$ and $x_{dt}$ 
is a non-linear equation for which there is no explicit solution in general. 
As discussed previously, in the $dt\to 0$ limit, one has $x_{dt}-x_0=O({dt}^{1/2})$ (which is true for instance for a Brownian motion when $f(x)=0$ and $g(x)=1$, 
and is checked  self-consistently in general). 
Writing $\bar x_0 = x_0 + \alpha (x_{dt}-x_0)$, we then expand~\eqref{eq:langevindiscalpha} in order to obtain $x_0 -x_{dt}$ up to order $O(dt)$ included.
One deduces
\begin{align}
  \label{eq:Deltax0}
  \fl\qquad\quad
  x_{dt}-x_0 &= 
\big[f (x_0 )+\eta _0 g (x_0 ) \big]dt
\nonumber
\\
&
\quad
+
\alpha  dt  (x_{dt}-x_0 )  \Big[f' (x_0 )+\eta _0 g' (x_0 )+\tfrac 12 \alpha  \eta _0  (x_{dt}-x_0 ) g'' (x_0 )\Big]
\end{align}
where we used $\eta_0=O({dt}^{-1/2})$. Solving for $x_{dt}$, 
one obtains, after expansion,
\begin{align}
  \label{eq:X1}
  \fl~
X_1(x_0,\eta_0)
=& 
\hspace*{-3mm}
x_0  
+dt\, f (x_0 )
+dt\, \eta_0\, g (x_0 )
+\alpha  dt^2  \big[f (x_0 )+\eta _0 g (x_0 ) \big]\big[f' (x_0 )+\eta _0 g' (x_0 ) \big]
\nonumber
\\
&
\hspace*{-4mm}
+\frac{1}{2} \alpha ^2 dt^3 \eta _0^2  
\Big\{
2 f (x_0 ) g' (x_0 )^2+\eta _0 g (x_0 )^2 g'' (x_0 )
\nonumber
\\[-2mm]
&
\hspace*{-6mm}
\phantom{+\frac{1}{2} \alpha ^2 dt^3 \eta _0^2 \{}
+
2 g (x_0 ) \Big[2 f' (x_0 ) g' (x_0 )+f (x_0 ) g'' (x_0 )+\eta _0 g' (x_0 )^2 \Big]
\Big\}
\;  , 
\end{align}
where we kept terms of high enough order in $\eta_0$ so as to ensure that the 
derivative w.r.t.~$\eta_0$ used in~\eqref{eq:Px1x0Pnoise} contains terms up to order $O(dt^2)$ included. This derivative reads
\begin{align}
\fl
\partial_{\eta_0}X_1(x_0,\eta_0)
=
dt\, g (x_0 )
\Big[
1
+
2 \alpha dt \eta_0\, g'& (x_0 )
+
\alpha dt  f'(x_0)
+
\alpha dt  f(x_0) \tfrac{g'(x_0)}{g(x_0)}
\nonumber
\\
&
+
\tfrac 32 dt^2 \alpha^2 \eta_0^2 \Big(  2 g'(x_0)^2+g(x_0)g''(x_0)  \Big)
\Big]
\; . 
\label{eq:derX1}
\end{align}
Inverting, we have
\begin{align}
\fl
    \frac
    1
    {\Big|\partial_{\eta_0}X_1\big(x_0,\eta_0\big)\Big|}
  =
    \frac
    1
    {|g(x_0) dt|}
    \Big[
    1-2 \alpha  dt &\eta _0\, g' (x_0 )
    -\alpha dt  f'(x_0)
    -\alpha  dt f (x_0 ) \tfrac{ g' (x_0 )}{g (x_0 )}
\nonumber
    \\[-3mm]
    &
    +\tfrac 12 dt^2 \alpha^2 \eta_0^2 \Big(  2 g'(x_0)^2-3g(x_0)g''(x_0)  \Big)
    \Big]
    \; . 
  \label{eq:invdhoX1}
\end{align}
In this expression, one can now use the substitution rule~(\ref{eq:itopresc_eta})
to derive
\begin{align}
\fl
\quad
    \frac
    1
    {\Big|\partial_{\eta_0}X_1\big(x_0,\eta_0\big)\Big|}
  =
    \frac
    1
    {|g(x_0) dt|}
    \Big[
    1-2 \alpha  dt \eta _0\, &g' (x_0 )
    -\alpha dt  f'(x_0)
    -\alpha  dt f (x_0 ) \tfrac{ g' (x_0 )}{g (x_0 )}
\nonumber
    \\[-6mm]
    &
    + D \alpha^2 dt \Big(  2 g'(x_0)^2-3g(x_0)g''(x_0)  \Big)
    \Big]
    \; . 
  \label{eq:invdhoX1b}
\end{align}
For later convenience, we prefer to express the numerator of the r.h.s.~in terms of $\bar x_0$ instead of $x_0$. We then 
utilise $x_0=\bar x_0-\alpha (x_{dt}-x_0)$ and we replace $x_{dt}-x_0$ by its expression deduced 
from~(\ref{eq:langevindiscalpha}). All in all, the only resulting non-trivial contribution to~(\ref{eq:invdhoX1b}) is
\begin{align}
dt \,\eta_0 g'(x_0)&=
dt \,\eta_0 g'\Big(\bar x_0-\alpha\,dt\,\big(g(\bar x_0)\eta_0 + f(\bar x_0)\big)\Big)
\nonumber\\
&=
dt \,\eta_0 g'(\bar x_0)-\alpha\,dt^2\eta_0^2\,g(\bar x_0)g''(\bar x_0)+o(dt)
\nonumber\\
&=
dt \,\eta_0 g'(\bar x_0)-2\alpha D\,dt\, g(\bar x_0)g''(\bar x_0)+o(dt)
\end{align}
and, finally, Eq.~\eqref{eq:invdhoX1b} becomes
\begin{align}
\fl
\quad
    \frac
    1
    {\Big|\partial_{\eta_0}X_1\big(x_0,\eta_0\big)\Big|}
  =
    \frac
    1
    {|g(x_0) dt|}
    \Big[
    1-2 \alpha  dt \eta _0\, & g' (\bar x_0 )
    -\alpha dt  f'(\bar x_0)
    -\alpha  dt f (\bar x_0 ) \tfrac{ g' (\bar x_0 )}{g (\bar x_0 )}
\nonumber
    \\[-5mm]
    &~
    + D \alpha^2 dt \Big(  2 g'(\bar x_0)^2+g(\bar x_0)g''(\bar x_0)  \Big)
    \Big]
    \; . 
  \label{eq:invdhoX1c}
\end{align}
At this point one would like to exponentiate this expression, an operation that 
has to be performed with care since $\eta_0 \, dt$ is of order $dt^{1/2}$.
Using the substitution rule~(\ref{eq:itopresc_eta}) as previously, one has
\begin{eqnarray}
  \ee^{A\eta_0 dt}&=&1+A\eta_0dt+DA^2dt+o(dt)
\; , \\
  \ee^{A\eta_0 dt-DA^2dt}&=&1+A\eta_0dt+o(dt)
 \; . 
  \label{eq:devexpWN}
\end{eqnarray}
These relations imply
\begin{align}
\fl
    \frac
    1
    {\Big|\partial_{\eta_0}X_1\big(x_0,\eta_0\big)\Big|}
  =
    \frac
    1
    {|g(x_0) dt|}
    \exp \Big[
    -2 \alpha  dt \eta _0\, g&' (\bar x_0 )
    -\alpha dt  f'(\bar x_0)
    -\alpha  dt f (\bar x_0 ) \tfrac{ g' (\bar x_0 )}{g (\bar x_0 )}
\nonumber
    \\[-5mm]
    & \hspace{-1cm}
    - D \alpha^2 dt \Big(  2 g'(\bar x_0)^2-g(\bar x_0)g''(\bar x_0)  \Big)
    \Big]
    \; . 
  \label{eq:invdhoX1d}
\end{align}

The other function that one needs to determine to compute the infinitesimal propagator~\eqref{eq:Px1x0Pnoise} is $H_0(x_0,x_{dt})$, defined by Eq.~\eqref{eq:defH0}, which is equivalent to Eq.~(\ref{eq:langevindiscalpha}) evaluated at $t=0$. 
After a simple rearrangement one finds
\begin{equation}
\eta_{0}=
\frac{1}{dt}
\frac
{x_{dt}-x_0-dt f (x_0 )}
{g(x_0)+\alpha   (x_{dt}-x_0 ) g' (x_0 )}
\equiv
H_0(x_0,x_{dt})
\; .
\label{eq:solH0notexpanded}
\end{equation}
In the denominator, one recognises an expansion around $\bar x_0$  [with the l.h.s.~$\eta_0$ evaluated up to $O(dt^0)$ included, so that $P_\text{noise}\big(H_0(x_0,x_{dt})\big)$ contains terms up to $O(dt)$, as seen from Eq.~(\ref{eq:dist})]. This yields
\begin{equation}
  \label{eq:H0}
H_0(x_0,x_{dt})
=
\frac
{\frac{x_{dt}-x_0}{dt}-f (\bar x_0 )}
{g(\bar x_0)}
\; . 
\end{equation}

\paragraphitalics{c Infinitesimal propagator}
\label{sec:infin-prop}
Coming back to Eq.~\eqref{eq:Px1x0Pnoise}, one obtains from Eqs.~(\ref{eq:dist}),  (\ref{eq:invdhoX1c}) and (\ref{eq:H0}) that
\begin{eqnarray}
\fl \mathbb P(x_{dt}|x_0) 
\stackrel\alpha 
=
\sqrt{\frac{dt^{-1}}{4\pi D}}
\frac
1
{|g(x_0)|}
\exp
\bigg\{
&
-\frac 12 \frac{dt}{2D} 
\bigg[\frac
{\frac{x_{dt}-x_0}{dt}-f (\bar x_0 )}
{g(\bar x_0)}\bigg]^2
  -\alpha dt  f'(\bar x_0)
\nonumber
\\
&   
-2 \alpha  dt 
\frac
{\frac{x_{dt}-x_0}{dt}-f (\bar x_0 )}
{g(\bar x_0)}
\, g' (\bar x_0 )
    -\alpha  dt f (\bar x_0 ) \tfrac{ g' (\bar x_0 )}{g (\bar x_0 )}
\nonumber
\\
&
    - D \alpha^2 dt \Big(  2 g'(\bar x_0)^2-g(\bar x_0)g''(\bar x_0)  \Big)
\bigg\}
\label{eq:preinftspropx0}
\end{eqnarray}
where the symbol $\stackrel\alpha= $ indicates that in the r.h.s.~$\bar x_0$ is the $\alpha$-discretised point.
Recognising a double-product to complete the square, one gets
\begin{small}
\begin{align}
\fl
  \mathbb P(x_{dt}|x_0) 
\stackrel\alpha= 
\sqrt{\frac{dt^{-1}}{4\pi D}}
\frac
1
{|g(x_0)|}
\exp\!
\bigg\{
\!
&
-\frac 12 \frac{dt}{2D} 
\bigg[\frac
{\frac{x_{dt}-x_0}{dt}-f (\bar x_0 )+4\alpha D\,g(\bar x_0)g'(\bar x_0)}
{g(\bar x_0)}\bigg]^2
   \! -\alpha dt  f'(\bar x_0)
\nonumber
\\
    &
    -\alpha  dt f (\bar x_0 ) \tfrac{ g' (\bar x_0 )}{g (\bar x_0 )}
    + D \alpha^2 dt \Big(  2 g'(\bar x_0)^2+g(\bar x_0)g''(\bar x_0)  \Big)
\!
\bigg\}
\:.
\label{eq:resinfinitesimalpropagx0}
\end{align}%
\end{small}%
The global prefactor $|g(x_0)|^{-1}$ in the infinitesimal 
propagator~\eqref{eq:resinfinitesimalpropagx0} can also be expressed in terms of $|g(\bar x_0)|^{-1}$.
For this, one starts again from $x_0=\bar x_0-\alpha (x_{dt}-x_0)$ and replaces $x_{dt}-x_0$ by its expression deduced from~(\ref{eq:langevindiscalpha}).
This leads to
\begin{align}
  \label{eq:gx0gx0bar}
\frac 1{|g(x_0)|}
=
\frac 1{|g(\bar x_0)|}
\bigg[
    1&+ \alpha  dt \eta _0\, g' (\bar x_0 )
    +\alpha  dt f (\bar x_0 ) \tfrac{ g' (\bar x_0 )}{g (\bar x_0 )}
\nonumber
    \\[-3mm]
    &
    + D \alpha^2 dt \Big(  2 g'(\bar x_0)^2-g(\bar x_0)g''(\bar x_0)  \Big)
\bigg]
\; . 
\end{align}
Exponentiating in the same way as we obtained~\eqref{eq:invdhoX1c}, 
\begin{align}
  \label{eq:gx0gx0barexp}
\frac 1{|g(x_0)|}
=
\frac 1{|g(\bar x_0)|}
\exp\! \bigg[
    1&+ \alpha  dt \eta _0\, g' (\bar x_0 )
    +\alpha  dt f (\bar x_0 ) \tfrac{ g' (\bar x_0 )}{g (\bar x_0 )}
\nonumber
    \\[-3mm]
    &
    + D \alpha^2 dt \Big(  g'(\bar x_0)^2-g(\bar x_0)g''(\bar x_0)  \Big)
\bigg]
\; . 
\!\!
\end{align}
Since this relation contains a term $\propto\eta _0$, once again one has to complete the square.
Coming back to Eq.~\eqref{eq:Px1x0Pnoise}, 
finally, many terms compensate and, instead of Eq.~(\ref{eq:resinfinitesimalpropagx0}), one obtains 
a simpler expression for the  infinitesimal propagator:
\begin{align}
\fl
\mathbb P(x_{dt}|x_0) 
\stackrel\alpha = 
\sqrt{\frac{dt^{-1}}{4\pi D}}
\frac
1
{|g(\bar x_0)|}
\nonumber\\
  \!\!\!\!\!\!\!\!
 \times \exp
\bigg\{
\!
-\frac 12 \frac{dt}{2D} 
\bigg[\frac
{\frac{x_{dt}-x_0}{dt}-f (\bar x_0 )+2\alpha D \,g(\bar x_0)g'(\bar x_0)}
{g(\bar x_0)}\bigg]^2
\!\!\!  - \alpha dt  f'(\bar x_0)
\!
\bigg\}
\:.
\label{eq:resinfinitesimalpropagx0b}
\end{align}

\paragraphitalics{d $\!$The continuous-time limit}
\label{sec:Discuss-prop}

The result~(\ref{eq:resinfinitesimalpropagx0b}) is well-known and can be derived in simpler ways~\cite{lau_state-dependent_2007,Aron_etal_2014,itami_universal_2017} that are reviewed in 
App.~\ref{sec:infin-prop-other}, and that do not use (or use in a different way) the equivalent substitution rules~(\ref{eq:itopresc}) or~(\ref{eq:itopresc_eta}). 
Having such different approaches leading to the same result is important in order to identify 
the conditions under which this substitution rule can be used;
we note in particular that we used this rule in Eqs.~(\ref{eq:invdhoX1b}), (\ref{eq:invdhoX1d}) and~(\ref{eq:gx0gx0barexp}) 
only in the prefactor of the exponential $P_\text{noise}(H_0)$ and not inside the exponential.
As discussed throughout Sec.~\ref{sec:stoch-calc-path}, such restriction on the condition under 
which the substitution rule~(\ref{eq:itopresc}) is valid proves to be crucial.

We can read from Eq.~(\ref{eq:resinfinitesimalpropagx0b}) the continuous-time limit~(\ref{eq:Feynman-path-int}): this yields the trajectory probability in the form~\eqref{eq:trajprobcont} with the so-called Onsager--Machlup action
\begin{align}
\label{eq:OMactionalpha}
\fl\qquad
  S[x(t)]
  \stackrel\alpha=
  \int_0^{\tf}\! dt\:
  \bigg\{
  \frac 12 \frac{1}{2D} 
  \bigg[\frac
  {\dt x-f (x)+2\alpha D \,g(x)g'(x)}
  {g(x)}\bigg]^2
  +
  \alpha  f'(x)
\bigg\}
\end{align}
where the arguments of the functions $f$ and $g$ are taken in $\alpha$-discretisation.
The associated normalisation prefactor reads
\begin{align}
  \mathcal J[x(t)]
  \stackrel\alpha=
  \prod_t 
\bigg\{
  \sqrt{\frac{dt^{-1}}{4\pi D}}
  \frac
  1
  {|g(\bar x_t)|}
\bigg\}
\; . 
\label{eq:Jacobialpha}
\end{align}
We emphasise (and this seems to have been little stressed in the literature) that it is essential to specify the discretisation point of the normalisation prefactor $\mathcal J[x(t)]$, since it can yield different contributions to the action for different discretisation conventions, as should be clear from 
Eq.~(\ref{eq:gx0gx0barexp}).
For instance, when proving the Fluctuation-Dissipation Theorem and the Fluctuation Theorem for Langevin equations with multiplicative noise~\cite{aron_dynamical_2016}, one has to 
take into account that reversing the time changes the discretisation from $\alpha$ to $1-\alpha$. 
This implies that when comparing the trajectory probability of a path and its time reversed, 
the discretisation of one of the normalisation prefactors has to be restored to $\alpha$ from $1-\alpha$, which induces terms similar to those 
in~(\ref{eq:gx0gx0barexp}) in the action without which the Fluctuation Theorem would not be verified.
%
%

We also note that Itami and Sasa have recently discussed in~\cite{itami_universal_2017} 
the consequences of choosing different $\alpha$-discretisations in the Langevin equation and in the action.

\medskip
\section{Stochastic calculus in the path integral action}
\label{sec:stoch-calc-path}

In general, the different actions $S[x(t)]$ that are studied in statistical mechanics (or in quantum field theory~\cite{zinn-justin_quantum_2002}) take the form of the time integral of a ``Lagrangian'': $S[x(t)]=\int_0^{\tf}dt\,\mathcal L(\dt x(t),x(t))$.
This is the case, for instance, of the action~\eqref{eq:OMactionalpha} that we derived in the previous Section and which corresponds to the $\alpha$-discretised Langevin equation~(\ref{eq:langevin_alpha}).
Since the trajectories $x(t)$ that verify the Langevin equation are not differentiable, it is natural to expect that  the Lagrangian $\mathcal L(\dt x(t),x(t))$ should be sensitive to the convention of $\alpha$-discretisation for its writing, and  that the differential transformations performed in the Lagrangian should incorporate terms akin to the stochastic ones $\propto (1-2\alpha)$ of the modified chain rule~(\ref{eq:chainrule_example}).

It is often assumed that the continuous-time chain rule~(\ref{eq:chainrule_example}) can be applied when manipulating the action (see for instance~\cite{tang_summing_2014}) or that the formulæ~(\ref{eq:langevin_alpha-to-alphabar})-(\ref{eq:langevin_alpha-to-alphabar_force}) describing the change of discretisation in the Langevin equation can be equally used.
In this Section, we show 
\begin{enumerate}
\item
that performing a change of discretisation in the Onsager--Machlup action is possible but completely wrong if one uses the relations~(\ref{eq:langevin_alpha-to-alphabar})-(\ref{eq:langevin_alpha-to-alphabar_force}); and 
\item
similarly, that non-linear changes of variables are allowed in the action but are also wrong if one applies the chain rule~(\ref{eq:chainrule_example}).
\end{enumerate}
In both cases, one arrives at inconsistencies when using the incorrect transformations to manipulate the action.
We describe the procedure which enables ones to
elucidate the origin of such apparent contradictions of stochastic calculus.
The moral of the story is that the substitution rule~(\ref{eq:itopresc}) alone is insufficient to understand the 
transformations of the action and that other transformations, that we derive, are needed. 

\subsection{From one discretisation to another}
\label{sec:from-discr-anoth}

\begin{figure}[t]
  \centering
  \includegraphics[width=.88\columnwidth]{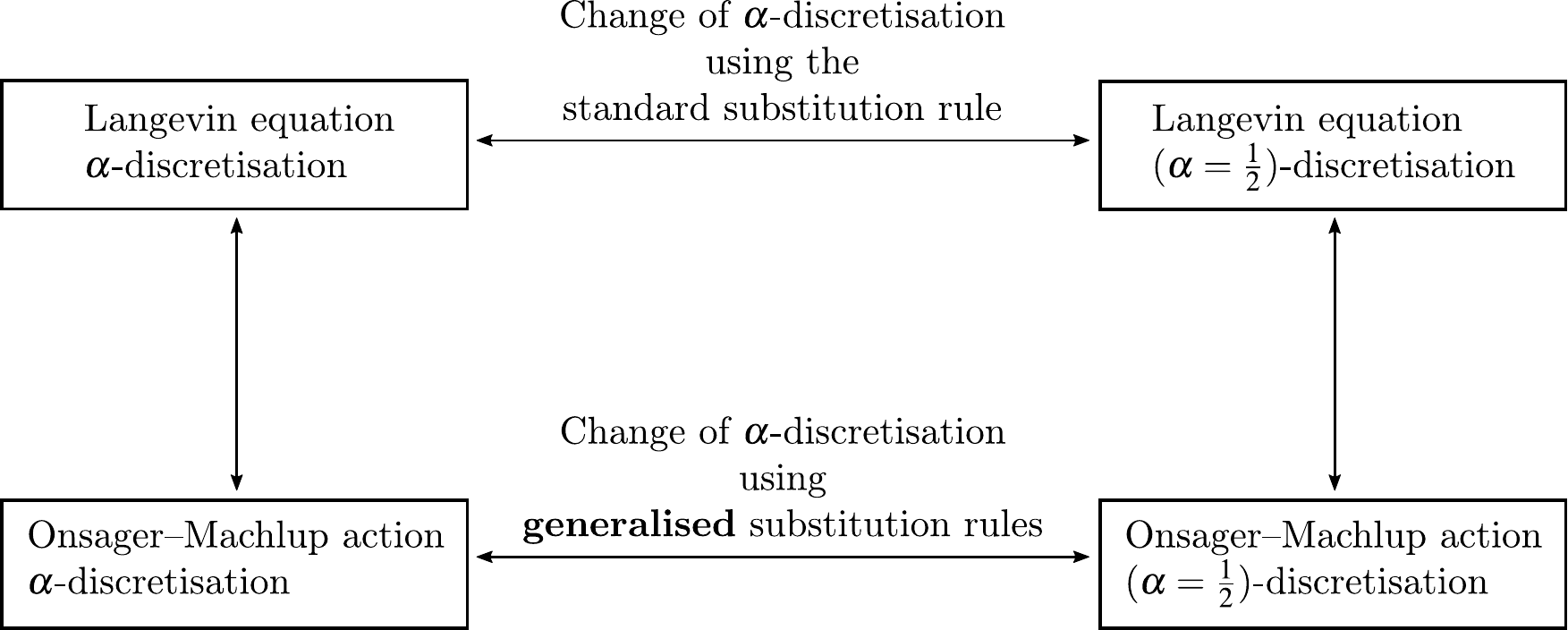}
  \caption{%
Schematic representation, for a change of discretisation, of the difference between the stochastic calculus performed in the Langevin equation and in the Onsager--Machlup action.
The $\alpha$-discretised Langevin equation~(\ref{eq:langevin_alpha}) can be transformed by use of the substitution rule~(\ref{eq:itopresc})
into a Stratonovich-discretised one ($\alpha=1/2$) given by Eqs.~(\ref{eq:langevin_Strato_force-alpha})-(\ref{eq:langevin_Strato_force-alpha_falpha}).
However, one \textbf{cannot} use such equations in the $\alpha$-discretised Onsager--Machlup action~(\ref{eq:OMactionalpha}) to get the correct Stratonovich-discretised action.
Instead, to go from one action to the other, one has to use the generalised substitution rules~(\ref{eq:Ito2b})-(\ref{eq:Ito6})
in discrete time for the infinitesimal propagator (once expanded in powers of $\Delta x $ and $dt$), or to rely on \emph{modified} substitution rules~(\ref{eq:Ito2MODIF})-(\ref{eq:Ito4MODIF}) inside the exponential of the propagator.
\label{fig:commutative-diag_change-of-discretisation}}
\end{figure}

In this subsection, we examine  the condition of validity of the transformation rules allowing one to go 
from an $\alpha$-discretised Langevin equation~(\ref{eq:langevin_alpha}) to an equivalent 
\mbox{$\bar\alpha$-discretised} Langevin equation~(\ref{eq:langevin_alpha-to-alphabar})-(\ref{eq:langevin_alpha-to-alphabar_force}). 
We focus on the transformation from a generic $\alpha$-discretisation to the Stratonovich one ($\alpha=1/2$), which is often performed for the reason that the stochastic chain rule~(\ref{eq:chainrule_example}) takes a simple form for $\alpha=1/2$ (yielding back the standard chain rule of differential calculus).
We show that,  although it seems natural to perform the same transformations in the action as at the Langevin level, such as changing discretisation through~(\ref{eq:langevin_alpha-to-alphabar})-(\ref{eq:langevin_alpha-to-alphabar_force}),  
the resulting action actually proves to be invalid (see Fig.~\ref{fig:commutative-diag_change-of-discretisation} for a schematic representation of the procedure). 
Finally, we identify the reason why the correct rules of calculus in the action are more complex than at the Langevin level, and we 
determine the  correct calculus to be used in the action that actually involves generalised substitution rules akin to~(\ref{eq:itopresc}).

\subsubsection{Direct change of discretisation in the action}
\label{sec:direct-change-discr}

The $\alpha$-discretised Langevin equation~(\ref{eq:langevin_alpha})
$
  \dt x \! \stackrel{\alpha}{=} f(x)+g(x)\,\eta
$  
is equivalent to the following Langevin equation in Stratonovich discretisation, with an $\alpha$-dependent force~$f_\alpha$
\begin{align}
  \label{eq:langevin_Strato_force-alpha}
  \dt x\ &
                 \stackrel{\text{\tiny{Strato}}}
                 =
                 \ f_\alpha(x)+g(x)\,\eta
                 \; , 
  \\
  \label{eq:langevin_Strato_force-alpha_falpha}
  f_\alpha(x) \ &
                  \stackrel{\phantom{\text{\tiny{Strato}}}}
                  =
                  f(x)+2D(\alpha-\tfrac 12) g'(x) g(x)
                  \; . 
\end{align}
This is seen, for instance, by coming back to the time-discrete definition (\ref{eq:langevindiscalpha})-(\ref{eq:langevindiscalpha_2}) 
of the $\alpha$-discretisation
and by working with the symmetric Stratonovich discretisation point (the superscript S indicates such choice of discretisation in what follows)
\begin{equation}
  \label{eq:def_x-strato}
  \bar x^\st_t = \tfrac 12 (x_{t+dt}+x_t)
  \; ,
\end{equation}
a procedure that we followed in Sec.~\ref{sec:chang-discr-while} for a generic change of discretisation: Eqs.~(\ref{eq:langevin_alpha-to-alphabar})-(\ref{eq:langevin_alpha-to-alphabar_force}) yield the result 
above, \emph{i.e.}~Eqs.~(\ref{eq:langevin_Strato_force-alpha})-(\ref{eq:langevin_Strato_force-alpha_falpha}) with 
a force $f_\alpha=f_{\alpha\to 1/2}$.

\subsubsection{Change of discretisation in the infinitesimal propagator}
\label{sec:change-discr-infin}

Since the $\alpha$-discretised Langevin equation~(\ref{eq:langevin_alpha}) and the Stratonovich 
one~(\ref{eq:langevin_Strato_force-alpha})-(\ref{eq:langevin_Strato_force-alpha_falpha}) are equivalent, they must possess equivalent infinitesimal propagators.
The change of discretisation in the infinitesimal propagator proves to be more involved than in the equation itself.

\paragraphitalics{a Expanding without throwing powers of $\Delta x$ out with the bathwater}
\label{sec:expand-with-throw}

We focus, without loss of generality, on the first time step $0 \curvearrowright dt$. 
The propagator~\eqref{eq:resinfinitesimalpropagx0b} in $\alpha$-discretisation is
\begin{align}
& 
\fl
  \mathbb P(x_{dt}|x_0) 
\stackrel\alpha= 
\frac
{\mathcal N}
{|g(\bar x_0)|}
\exp
  \bigg\{
  \!\!
  -\frac 12 \frac{dt}{2D} 
  \bigg[  \frac
  {\frac{x_{dt}-x_0}{dt}-f (\bar x_0 )+2\alpha D \,g(\bar x_0)g'(\bar x_0)}
  {g(\bar x_0)}\bigg]^2
\nonumber
\!\!\!  -\alpha dt  f'(\bar x_0)
  \!
  \bigg\}
  \\
& \text{with} \quad  {\mathcal N} \equiv \sqrt{\frac{dt^{-1}}{4\pi D}}\;.
\label{eq:resinfinitesimalpropagx0b_2}
\end{align}
The aim is to determine an equivalent propagator in terms of the Stratonovich mid-point $\bar x^\st_0 = \tfrac 12 (x_{dt}+x_0)$.
We expand~(\ref{eq:resinfinitesimalpropagx0b_2}) in powers of $\Delta x\equiv x_{dt}-x_0$, using
\begin{equation}
  \label{eq:x_alpha-to-strato_t0}
  \bar x_0
  = \bar x_0^\st +(\alpha-\tfrac 12) \Delta x
\end{equation}
and keeping all terms of order $dt^0$ inside the exponential (note that they \emph{define} the Gaussian weight), while putting all terms of order $O(dt^{1/2})$ and $O(dt)$ in a prefactor of this weight.
In this procedure, one should remember that $\Delta x=O(dt^{1/2})$. This crucially implies that, in the exponential, the expansion of the term
\begin{equation}
  \label{eq:difficult-term-expansion}
  -\frac 12 \frac{1}{2D\,dt} 
  \bigg[\frac
  {\Delta x}
  {g\big(\bar x_0^\st +(\alpha-\tfrac 12) \Delta x\big)}\bigg]^2
\end{equation}
generates terms of order $O(dt^{1/2})$ and  $O(dt)$ which are proportional to $dt^{-1} \Delta x^3$ and $dt^{-1} \Delta x^4$.
Expanding then the exponential, one gets terms up to $dt^{-2} \Delta x^6$. Explicitly, the result is
\begin{small}
\begin{align}
\fl
 \frac{
    \mathbb P(x_{dt}|x_0)
  }
  {
  \frac
  {\mathcal N}
  {|g(\xs_0)|}
  \ee^{
  \!
  -\frac 12 \frac{dt}{2D} 
  \!\!
  \textnormal{
  $\big(\frac{\Delta x}{dt}\big)^2$
  }
  \!\!\!\!\big/ {g(\xs_0)^2}
  }
  }
&\stackrel{\st}=
  \nonumber
  \\ 
& \hspace*{-25mm}
  \!\!\!1
    +
    \bigg[
    \frac{f(\xs_0)}{2 D g(\xs_0)^2}+\frac{(2-8 \alpha ) g'(\xs_0)}{4 g(\xs_0)}
    \bigg] \Delta x
    +
    \bigg[
    \frac{(2 \alpha -1) g'(\xs_0)}{4 D g(\xs_0)^3}
    \bigg] \Delta x^3 dt^{-1}
    \nonumber
  \\
& \hspace*{-25mm}
  +
  \bigg[
    -\alpha    \Big(\alpha  D g'(\xs_0)^2+f'(\xs_0)\Big)
    +\frac{\alpha    f(\xs_0) g'(\xs_0)}{g(\xs_0)}
    -\frac{  f(\xs_0)^2}{4 D g(\xs_0)^2}
    \bigg] dt
    \nonumber
  \\[2mm]
& \hspace*{-25mm}
    +
    \bigg[
    \frac{(-12 \alpha ^2+8 \alpha -1) g''(\xs_0)}{8 g(\xs_0)}
    +\frac{(14 \alpha ^2-8 \alpha +1) D g'(\xs_0)^2+(2 \alpha -1) f'(\xs_0)}{4 D g(\xs_0)^2}
    \nonumber
  \\
& \hspace*{-25mm}
    \phantom{    \qquad\qquad    \bigg[    \frac{(-12 \alpha ^2+8 \alpha -1) g''(\xs_0)}{8 g(\xs_0)}  }
    +\frac{(3-8 \alpha ) f(\xs_0) g'(\xs_0)}{4 D g(\xs_0)^3}
    +\frac{f(\xs_0)^2}{8 D^2 g(\xs_0)^4}
    \bigg] \Delta x^2
    \nonumber
  \\[2mm]
& \hspace*{-25mm}
  +
  \bigg[
  \frac{(1-2 \alpha )^2 g''(\xs_0)}{16  D g(\xs_0)^3}
  -\frac{(28 \alpha ^2-24 \alpha +5) g'(\xs_0)^2}{16  D g(\xs_0)^4}
  +\frac{(2 \alpha -1) f(\xs_0) g'(\xs_0)}{8  D^2 g(\xs_0)^5}
  \bigg] \Delta x^4\ dt^{-1}
    \nonumber
  \\[2mm]
& \hspace*{-25mm}
  +
  \frac{(1-2 \alpha )^2 g'(\xs_0)^2}{32 D^2 g(\xs_0)^6}
  \ \Delta x^6\ dt^{-2}
  \; .
\label{eq:resdev}
\end{align}%
\end{small}%
Note that we also changed the discretisation of the 
normalisation prefactor from $1/|g(\bar x_0)|$ to $1/|g(\xs_0)|$ 
using a relation similar to~(\ref{eq:gx0gx0bar}).
The symbol $\stackrel\st=$ indicates that the r.h.s.~is evaluated in the Stratonovich discretisation.

\paragraphitalics{b Comparison to the propagator arising from changing discretisation at the Langevin level}
\label{sec:comp-prop-aris}

We would like to compare this result to that of the commutative 
procedure depicted in Fig.~\ref{fig:commutative-diag_change-of-discretisation}, namely,
\begin{enumerate}
\item 
transform the original $\alpha$-discretised Langevin equation into the Stratonovich-discretised one~(\ref{eq:langevin_Strato_force-alpha}) 
which includes an $\alpha$-dependent force $f_\alpha(x)$ given by~(\ref{eq:langevin_Strato_force-alpha_falpha}); and 
\item 
follow the same procedure as previously done 
to get the corresponding propagator, that we denote $\mathbb P^\st_{f_\alpha}$. 
\end{enumerate}
The result is, of course, directly read from Eq.~(\ref{eq:resinfinitesimalpropagx0b_2}), where $\alpha$ is first replaced by 
$1/2$ (and hence $\bar x_0$ by $\xs_0$), and then $f$ is replaced by $f_\alpha$; this yields
\begin{small}
\begin{align}
\fl\quad
  \mathbb P^\st_{f_\alpha}(x_{dt}|x_0) 
\stackrel\st= 
\frac
{\mathcal N}
{|g(\xs_0)|}
\exp
  \bigg\{
  \!\!
  -\frac 12 \frac{dt}{2D} 
  \bigg[\frac
  {\frac{\Delta x}{dt}-f_\alpha(\xs_0 )+ D \,g(\xs_0)g'(\xs_0)}
  {g(\xs_0)}\bigg]^2
  \!\!\!\!
  -\frac 12 dt  f_\alpha'(\xs_0)
  \bigg\}
  \; . 
\label{eq:resinfinitesimalpropagx0b_2_commut}
\end{align}%
\end{small}%
By consistency, this propagator should be equal to the result~(\ref{eq:resinfinitesimalpropagx0b_2}), in the small $dt$ limit. 
To check whether this is the case, we follow the same procedure as the one leading to Eq.~(\ref{eq:resdev}) from 
Eq.~(\ref{eq:resinfinitesimalpropagx0b_2}), that is to say, we expand in powers of $\Delta x$ and $dt$, and 
we replace $f_\alpha$ by its explicit expression in terms of $f$, $g$ and $\alpha$, to obtain
\begin{small}
\begin{align}
\fl  \frac{
\mathbb P^\st_{f_\alpha}(x_{dt}|x_0) 
  }
  {
  \frac
  {\mathcal N}
  {|g(\xs_0)|}
  {\ee}^{
  \!\!
  -\frac 12 \frac{dt}{2D} 
  \!\!
  \textnormal{
  $\big(\frac{\Delta x}{dt}\big)^2$
  }
  \!\!\big/ {g(\xs_0)^2}
  }
  }
&\stackrel\st= 
  \nonumber
  \\ 
& \hspace*{-20mm}
  \!\!\!1
    +
    \bigg[
    \frac{f(\xs_0)}{2 D g(\xs_0)^2}+\frac{(\alpha -1) g'(\xs_0)}{g(\xs_0)}
    \bigg] \Delta x
    \nonumber
  \\[-0mm]
  & \hspace*{-20mm}
    +
    \bigg[
-\frac{f(\xs_0)^2}{2 D g(\xs_0)^2}
-f'(\xs_0)-\frac{2 (\alpha -1) f(\xs_0) g'(\xs_0)}{g(\xs_0)}
\nonumber
\\[-2mm]
& \hspace*{-10mm}
\quad\quad
+
D 
  \Big\{
  (1-2 \alpha ) g(\xs_0) g''(\xs_0)+(-2 (\alpha -1) \alpha -1) g'(\xs_0)^2
  \Big\}
    \bigg]
    \frac 12
    dt
\nonumber
  \\
& \hspace*{-20mm}
    +
    \frac{(2 (\alpha -1) D g(\xs_0) g'(\xs_0)+f(\xs_0))^2}{8 D^2 g(\xs_0)^4}
    \
    \Delta x^2
    \; . 
\label{eq:resdevcommut}
\end{align}%
\end{small}%
The result is clearly different from the one in Eq.~(\ref{eq:resdev}), while one expects 
  $ \mathbb P(x_{dt}|x_0)= \mathbb P^\st_{f_\alpha}(x_{dt}|x_0)$ 
because these two propagators correspond to the same Langevin equation.
In particular,  the maximum power of $\Delta x$ for  $\mathbb P^\st_{f_\alpha}(x_{dt}|x_0)$ in Eq.~(\ref{eq:resdevcommut}) is $\Delta x^2$
while it is $\Delta x^6$ in Eq.~(\ref{eq:resdev}) for  $ \mathbb P(x_{dt}|x_0)$.

Note that if one takes for $g(x)$ a constant function $g$, the two propagators are still different, as checked by direct inspection (unless $\alpha=1/2$, as it should  
because then there is no change of discretisation and the two computations are identical).
The simple case of additive noise, thus, also requires a peculiar attention.

\paragraphitalics{c Appropriate substitution rules to render the two approaches compatible}
\label{sec:prescriptions}

As discussed in Sec.~\ref{sec:chain-rule}, the Itō prescription amounts to using the substitution rule
\begin{equation}
  \label{eq:Ito2_action}
  \Delta x^2 \mapsto 2D g(x)^2 dt  \quad \textnormal{as $dt\to 0$}\;,
\end{equation}
where on the r.h.s., the argument $x$ of $g(x)$ can be taken at any discretisation point, at minimal order in $dt$.
We have seen in paragraph~\ref{sec:infin-prop}\hyperref[{sec:infin-prop}]{.c} that the use of such prescription is justified as long as it is performed outside the exponential, 
for the determination of the infinitesimal propagator. 

\medskip
Therefore, in order to recover  from~(\ref{eq:resdev})  the simpler result~(\ref{eq:resdevcommut}) for the propagator, a natural possibility is to look for ``generalised substitution rules'' akin to~(\ref{eq:Ito2_action}), but now for terms of 
the form $\Delta x^n dt^m$ with $n,m$ chosen so that $\Delta x^n dt^m$ is typically of order $O(dt^{1/2})$ or $O(dt)$.
One finds by direct computation that, to guarantee that~(\ref{eq:resdev}) becomes~(\ref{eq:resdevcommut}), there is a \emph{unique} prescription to replace the terms $\Delta x^n dt^m$ by standard infinitesimals of the natural form $\text{C}^\text{st}\times (2Dg(x)^2)^{n/2} dt$ when $n$ is even and  $\text{C}^\text{st}\times\Delta x\, (2Dg(x)^2)^{(n-1)/2}$ when $n$ is odd. It reads\\
\begin{minipage}{1.0\linewidth}
\begin{align}
  \label{eq:Ito2b}
  \Delta x^2 &= 2D g(x)^2 \,dt
  \; , 
  \\
  \label{eq:Ito3}
  \Delta x^3 \,dt^{-1} &= 3 \ \Delta x \ 2D g(x)^2 
 \; ,  \\
  \label{eq:Ito4}
  \Delta x^4 \,dt^{-1} &= 3 \  \big(2D g(x)^2\big)^2 \,dt
  \; , \\
  \label{eq:Ito6}
  \Delta x^6 \,dt^{-2} &= 15 \ \big(2D g(x)^2\big)^3 \,dt
  \; . 
\end{align}
\end{minipage}
\smallskip

A justification of these generalised substitution rules,
to be understood in a precise~$L^2$ sense, 
is presented in App.~\ref{sec:just-gener-it=o}. It is similar in spirit to the usual mathematical definition of the 
first rule (the usual Itō prescription~(\ref{eq:itopresc})), the  $L^2$ definition of which is also recalled in this Appendix.

\paragraphitalics{d Discussion and comparison to a naive continuous-time computation}
\label{sec:discussion_changeofdiscretisation}

In Sec.~\ref{sec:direct-change-discr} we showed that the change of discretisation at the Langevin equation level requires  the use of the standard substitution rule~(\ref{eq:itopresc}) (the Itō prescription). This transformation follows the upper branch in Fig.~\ref{fig:commutative-diag_change-of-discretisation}.
In Sec.~\ref{sec:comp-prop-aris}\hyperref[{sec:comp-prop-aris}]{.c} we proved that the change of discretisation at the Onsager--Machlup level (for the infinitesimal propagator) 
requires a full set of generalised substitution rules, given by the relations~(\ref{eq:Ito2b})-(\ref{eq:Ito6}), that include the Itō prescription~(\ref{eq:itopresc})
but extend it with transformation rules for three other infinitesimals.
This transformation follows the 
lower branch in Fig.~\ref{fig:commutative-diag_change-of-discretisation}. Therefore, the paths along the upper and lower branches 
should be followed using procedures that involve a different set of substitution rules.

The key point that explains the discrepancy between the two approaches
is that when one changes the discretisation in the action, the term which is quadratic in $\Delta x$ (see~(\ref{eq:difficult-term-expansion})) 
transforms in a non-trivial way and
contributes to a higher order in powers of $\Delta x$ than when one changes the discretisation in the 
Langevin equation (as done in subsec.~\ref{sec:direct-change-discr}). 
Technically, the presence of a square $(\Delta x/dt)^2$ divided by the noise amplitude in the infinitesimal propagator 
implies that, when keeping terms of order $O(dt^{1/2})$ and $O(dt)$, 
higher powers of $\Delta x$ are generated, as observed in~(\ref{eq:resdev}). 

An instructive observation  is to draw a comparison between the Stratonovich-discretised continuous-time action corresponding to~(\ref{eq:resinfinitesimalpropagx0b_2_commut})
\begin{align}
\fl\qquad\quad
S^\st_{f_\alpha}[x(t)]
\stackrel{\st}= 
\frac 12 \int_0^{\tf}
dt\:
  \bigg\{
\frac{1}{2D} 
  \bigg[\frac
  {{\dt}x-f_\alpha(x)+ D \,g(x)g'(x)}
  {g(x)}\bigg]^2
  \!\!\!\!
  - dt  f_\alpha'(x)
  \bigg\}
\label{eq:actionalphastratochangeinLang}
\end{align}
and the result of a naive computation.
First, one notes that both the right-down and the down-right branches of the commutative diagram represented in Fig.~\ref{fig:commutative-diag_change-of-discretisation} agree with the same result~(\ref{eq:actionalphastratochangeinLang}), together with the corresponding prefactor $\prod_t{|g(\xs_t)|}$; this is true
%
%
provided one uses the generalised substitution rules~(\ref{eq:Ito2b})-(\ref{eq:Ito6}).
Another --~naive~-- approach consists in attempting to arrive at this result by changing the  discretisation 
directly in the time-continuous action, with the following procedure:
\begin{enumerate}
\item start from the continuous-time $\alpha$-discretised action~\eqref{eq:OMactionalpha};
\item use the
  rules~(\ref{eq:langevin_alpha-to-alphabar})-(\ref{eq:langevin_alpha-to-alphabar_force}) for the change of discretisation in the Langevin equation;
\item change the discretisation of the normalisation prefactor~\eqref{eq:Jacobialpha} from $\alpha$ to Stratonovich, using a relation similar
to~(\ref{eq:gx0gx0barexp})%
\footnote[2]{%
The relation~(\ref{eq:gx0gx0barexp}) allows one to change the discretisation of the prefactor~$\mathcal J[x(t)]$ from the Itō one 
($\alpha=0$) to the $\alpha$ one, but is easily adapted to change from $\alpha$ to Stratonovich ($\alpha=1/2$); see~Eq.~\eqref{eq:contribprefactorchdiscr}. 
}%
.
\end{enumerate}
However, as detailed in App.~\ref{sec:inconsistency_action_ch-discretis}, the result of this procedure is different from~(\ref{eq:actionalphastratochangeinLang}) and is thus \emph{incorrect}.
The reason lies in the fact that the
rules~(\ref{eq:langevin_alpha-to-alphabar})-(\ref{eq:langevin_alpha-to-alphabar_force})
for the change of discretisation in the Langevin equation do not involve substitution rules of high enough order in $\Delta x$: they 
disregard essential terms contributing to the expansion~(\ref{eq:resdev}) that are crucial to arrive at the final correct 
propagator~(\ref{eq:resinfinitesimalpropagx0b_2_commut}) (or, equivalently, to recover the correct action~(\ref{eq:actionalphastratochangeinLang}) 
with its associated Stratonovich-discretised normalisation prefactor). 
This confirms  
that the sole standard substitution rule~(\ref{eq:itopresc}) is not sufficient
to handle successfully the path integral representation of the stochastic process, and that the generalised substitution rules (\ref{eq:Ito2b})-(\ref{eq:Ito6}) that we 
propose have to be used instead.

\subsection{Non-linear transformation} 
\label{sec:non-line-transf}


A similar apparent contradiction occurs when one attempts to use the chain rule~(\ref{eq:chainrule_example}) in the action, instead of restricting its use to the Langevin level.
Such inconsistency was observed for non-linear field transformations for the MSRJD action in App.~E of~\cite{aron_dynamical_2014_arxiv1}%
\footnote{%
Note that this Appendix is found only in the arXiv v1 preprint version.
}%
.
In App.~\ref{sec:inconsistency_action_chain-rule} of the present article, we translate this computation to the 
case of the Onsager--Machlup action, and the result is the same: using the chain rule~(\ref{eq:chainrule_example}) in the action brings an inconsistency when changing variables.
Related issues have been observed in the context of quantum field theory~\cite{gervais_point_1976,Sa77,LaRoTi79,Tirapegui82,AlDa90,ApOr96}.
In this subsection, we examine the origin of this paradox. 
We show that it is again  due to an invalid use of the Itō substitution rule~(\ref{eq:itopresc}) in the action, and we provide the correct treatment of 
non-linear transformations in the action, working with the infinitesimal propagator.
We also propose a modified chain rule that can be used in continuous time inside the action.

\subsubsection{Non-linear transformation in the Langevin equation}
\label{sec:non-line-transf-1}

\begin{figure}[t]
  \centering
  \includegraphics[width=.88\columnwidth]{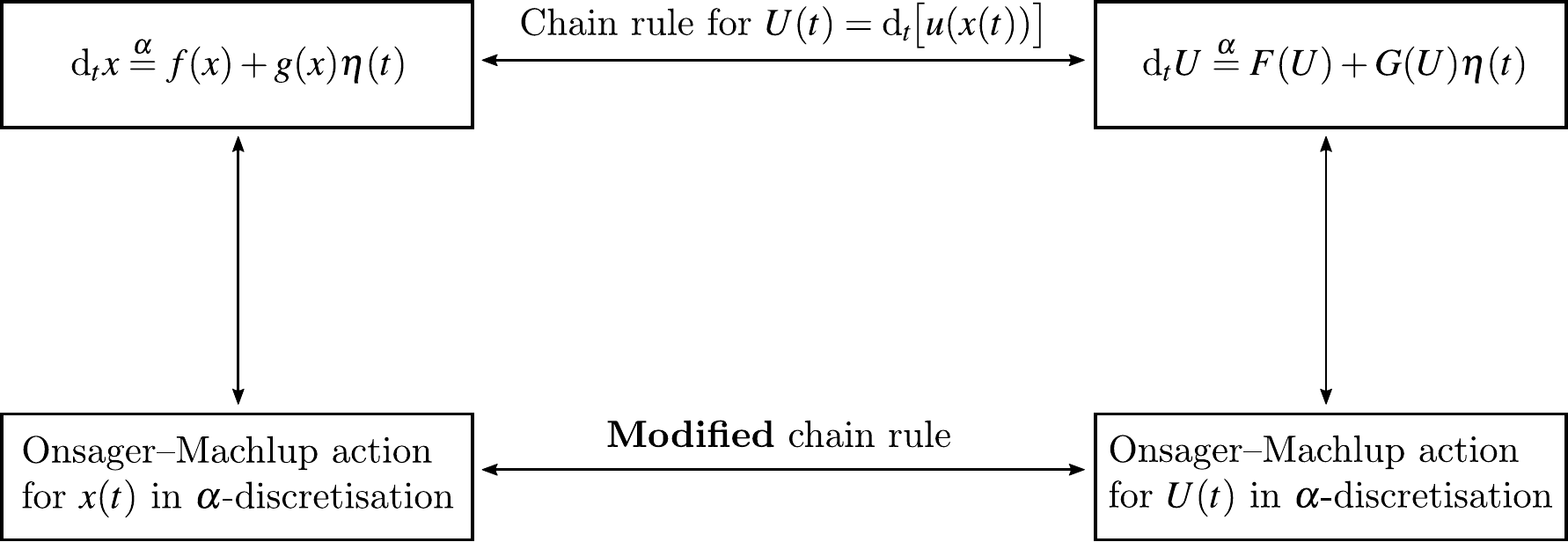}
  \caption{%
Schematic representation, for a non-linear (bijective) change of variables $x\mapsto u(x)$, of the difference between the stochastic calculus performed in the Langevin equation and in the Onsager--Machlup action.
The $\alpha$-discretised Langevin equation~(\ref{eq:langevin_alpha}) can be transformed exploiting the chain rule~(\ref{eq:chainrule_example}) into a Langevin equation~(\ref{eq:langevin_alpha_u}) for $U(t)=u(x(t))$.
To this equation corresponds a Onsager--Machlup action, equivalent to the infinitesimal propagator~(\ref{eq:resinfinitesimalpropagx0bU}).
However, one \textbf{cannot} use such chain rule in the $\alpha$-discretised Onsager--Machlup action~(\ref{eq:OMactionalpha}) to derive the correct action of the process $U(t)$, as explained in App.~\ref{sec:inconsistency_action_chain-rule}.
To go from one action to the other, one  has to use, instead, the generalised substitution rules~(\ref{eq:Ito2b})-(\ref{eq:Ito6}) in discrete time after expanding the action [or equivalently~(\ref{eq:Ito2MODIF})-(\ref{eq:Ito4MODIF}) inside the discretised action], or to rely on a \emph{modified} chain rule for the time-continuous process as discussed in subsec.~\ref{sec:modified-chain-rule}.
In the text, for simplicity, the lower branch of this commutative diagram is performed from right to left.
\label{fig:commutative-diag_non-lin-variable-change}%
}
\end{figure}

We start from the $\alpha$-discretised Langevin equation~(\ref{eq:langevin_alpha})
and consider an increasing $C^1$ function $u(x)$ which is a bijection and is used as a non-linear change of variables.
The chain rule~(\ref{eq:chainrule_example})
implies that the function $U(t)=u(x(t))$ satisfies an $\alpha$-discretised Langevin equation
\begin{equation}
  \label{eq:langevin_alpha_u}
  \dt U\ \stackrel\alpha=\
    u'f  + (1-2\alpha) D\,g^2u''
 +
    u'g\, \eta
\;.
\end{equation}
This writing is a shortcut for the Langevin equation with a force $F$ and a noise amplitude~$G$
\begin{align}
  \label{eq:langevin_alpha_uU}
 \dt  U(t)\ \stackrel\alpha
  =&\
  \overbrace{
    u'(x_U(t))\, f(x_U(t)) + (1-2\alpha) D\,g(x_U(t))^2u''(x_U(t))
  }^{
    \equiv F(U(t))
  }
\nonumber
\\
&\  \ + \
  \underbrace{
    u'(x_U(t))\,g(x_U(t))
  }_{
    \equiv G(U(t))
  }
  \eta(t)
  \\
\;\;
  x_U(t)
\:
  = &\
      u^{-1}(U(t))  \qquad \big(\text{\emph{i.e.}}~u(x_U(t))= U(t)\big)
\;.
\end{align}
Our aim is to compare different procedures represented on the commutative diagram of Fig.~\ref{fig:commutative-diag_non-lin-variable-change}.
Concretely we take the following two paths.
\begin{enumerate}
\item The down path (on the left) that starts from the
  $\alpha$-discretised Langevin equation~(\ref{eq:langevin_alpha}) and
  arrives at the Onsager--Machlup action on $x(t)$ given
  by the expression in~(\ref{eq:OMactionalpha}), which, together with its associated
  normalisation prefactor~(\ref{eq:Jacobialpha}), is equivalent to the infinitesimal
  propagator~(\ref{eq:resinfinitesimalpropagx0b}).
\item The right-down-left path. It starts from the Langevin
  equation~(\ref{eq:langevin_alpha_u}), goes next to its corresponding
  Onsager--Machlup representation and, finally, through the application of
  rules that we still need to find, this path performs a non-linear transformation on the
  Onsager--Machlup action on $U(t)$ that should take it to the one on $x(t)$.
\end{enumerate}
We first analyse these procedures at the infinitesimal propagator level.

\bigskip
\subsubsection{Direct determination of the propagator}
\label{sec:direct-determ-prop}

As in subsec.~\ref{sec:change-discr-infin}, we perform the comparison by keeping 
only the quadratic in $\Delta x$ contribution to the Gaussian weight in the exponential, and by expanding the rest in front of this weight.
The propagator~(\ref{eq:resinfinitesimalpropagx0b}) associated to the Langevin equation~(\ref{eq:langevin_alpha}) reads
\begin{align}
\fl\quad
  \mathbb P(x_{dt}|x_0) 
\stackrel\alpha= 
\frac
{\mathcal N}
{|g(\bar x_0)|}
  \:
&  \ee^{
  -\frac 12 \frac{dt}{2D} 
  \!\!
  \textnormal{
  $\big(\frac{\Delta x}{dt}\big)^2$
  }
  \!\!\!\big/ {g(\bar x_0)^2}
  }
\nonumber
\\
&
\times
  \bigg\{
  1-dt \alpha  f'\left(\bar{x}_0\right)+\frac{
   f\left(\bar{x}_0\right)-2 D \alpha  g\left(\bar{x}_0\right)
  g'\left(\bar{x}_0\right)}{2 D g\left(\bar{x}_0\right){}^2} \Delta x
  \bigg\}
  \; . 
\label{eq:resinfinitesimalpropag_nlin_dev}
\end{align}
In this expansion, we have already used the standard substitution rule~(\ref{eq:itopresc}) to reexpress~$\Delta x^2$.

\bigskip
\subsubsection{Indirect path: passing through the propagator for $U(t)$}
\label{sec:indir-path:-pass}

Corresponding to the Langevin equation~(\ref{eq:langevin_alpha_uU}) for $U(t)$, 
one can write from~(\ref{eq:resinfinitesimalpropagx0b}) the propagator
\begin{small}
\begin{align}
\fl\quad
  \mathbb P_U(U_{dt}|U_0) 
\stackrel\alpha= 
\frac
{\mathcal N}
{|G(\bar U_0)|}
\exp
\bigg\{
&-\frac 12 \frac{dt}{2D} 
\bigg[\frac
{\frac{U_{dt}-U_0}{dt}-F (\bar U_0 )+2\alpha D \,G(\bar U_0)G'(\bar U_0)}
{G(\bar U_0)}\bigg]^2
\nonumber
\\
&
\qquad\qquad\qquad\qquad\qquad\qquad
    -\alpha dt  F'(\bar U_0)
\bigg\}
\; . 
\label{eq:resinfinitesimalpropagx0bU}
\end{align}
\end{small}%
Since the two Langevin equations~(\ref{eq:langevin_alpha}) and~(\ref{eq:langevin_alpha_uU}) are equivalent, 
this propagator has to be equivalent to~(\ref{eq:resinfinitesimalpropag_nlin_dev}).
As remarked in the literature in the stochastic~\cite{Langouche81,aron_dynamical_2014_arxiv1} and the quantum mechanical~\cite{gervais_point_1976,Sa77,LaRoTi79,Tirapegui82,AlDa90,ApOr96} contexts, the application of the
chain rule 
does \emph{not} yield back~(\ref{eq:resinfinitesimalpropag_nlin_dev}) or~(\ref{eq:resinfinitesimalpropagx0b}).
The computation describing this inconsistency for our Onsager--Machlup action of interest is recalled for 
completeness in App.~\ref{sec:inconsistency_action_chain-rule} 

\medskip

The idea to examine the origin of this inconsistency, as done previously for the change of discretisation, is to 
treat the ``dangerous'' term of the propagator $\big[\frac{U_{dt}-U_0}{dt \, G(\bar U_0)}\big]^2$ in a safe way, by 
expanding the propagator and putting all terms in prefactor, apart from the quadratic part defining the Gaussian weight itself.
To set up the expansion, one uses that
\begin{align}
  \bar U_0 &= (1-\alpha)u(x_0) + \alpha\, u(x_{dt})
  \; , 
\label{eq:Ubar0}
  \\
  x_0 &= \bar x_0 - \alpha\, \Delta x
  \; , 
\label{eq:X0}
  \\
  x_{dt} &= \bar x_0 + (1-\alpha) \Delta x
  \; , 
\label{eq:Xdt}
\end{align}
and one expands in powers of $\Delta x$, keeping in mind that this quantity is $O(dt^{1/2})$.
The change of variables in the (conditional) probability 
\begin{equation}
  \label{eq:change_var}
  \mathbb P_U(U_{dt}|U_0)  \ {u'(x_{dt})}
  =
  \mathbb P(x_{dt}|x_0) 
\end{equation}
is also needed, where in $u'(x_{dt})$ one uses~(\ref{eq:Xdt}).
After a tedious computation (where  the substitution rule~(\ref{eq:Ito2_action}) for $\Delta x^2$ is employed though only in the prefactor), 
the result is that the propagator~$  \mathbb P(x_{dt}|x_0) $ obtained from~(\ref{eq:change_var}), with $\mathbb P_U$  read from~(\ref{eq:resinfinitesimalpropagx0bU}), is
\begin{small}
\begin{align}
\fl
    \frac
    {
    \mathbb P(x_{dt}|x_0)
    }
    {
    \frac
    {\mathcal N}
    {|g(\bar x_0)|}
    \ee^{
    \!
    -\frac 12 \frac{dt}{2D} 
    \!\!
    \textnormal{
    $\big(\frac{\Delta x}{dt}\big)^2$
    }
    \!\!\!\!\big/ {g(\bar x_0)^2}
    }
    }
  &
\nonumber
  \\
&
  \hspace*{-32mm}
  \stackrel\alpha=
  1
    +
    \bigg[
    \frac{f(\bar{x}_0)}{2 D g(\bar{x}_0){}^2}-\frac{\alpha  g'(\bar{x}_0)}{g(\bar{x}_0)}-\frac{3 (-1+2 \alpha ) u''(\bar{x}_0)}{2 u'(\bar{x}_0)}
    \bigg] \Delta x
    \nonumber
  \\[2mm]
  &
  \hspace*{-30mm}
    +
    \bigg[
    \frac{3  D \alpha  (-2+3 \alpha ) g(\bar{x}_0) g'(\bar{x}_0) u''(\bar{x}_0)}{u'(\bar{x}_0)}
    -
    \frac{ (2 \alpha  f'(\bar{x}_0) u'(\bar{x}_0)+3 (-1+2 \alpha ) f(\bar{x}_0) u''(\bar{x}_0))}{2 u'(\bar{x}_0)}+
    \nonumber
  \\
  &
  \hspace*{-20mm}
    +
    \frac{ D g(\bar{x}_0){}^2 (3 (1-6 \alpha +6 \alpha ^2) u''(\bar{x}_0){}^2+2 (1-3 \alpha +3 \alpha ^2) u'(\bar{x}_0) u^{(3)}(\bar{x}_0))}{2 u'(\bar{x}_0){}^2}
    \bigg]\, dt
    \nonumber
  \\
  &
    \hspace*{-30mm}
    +
    \frac{(-1+2 \alpha ) u''(\bar{x}_0)}{4 dt D g(\bar{x}_0){}^2 u'(\bar{x}_0)}    
    \:\Delta x^3
    %
\nonumber
  \\[2mm]
  &
  \hspace*{-30mm}
    +
    \bigg[
    \frac{(-1+2 \alpha ) f(\bar{x}_0) u''(\bar{x}_0)}{8 dt D^2 g(\bar{x}_0){}^4 u'(\bar{x}_0)}
    -
    \frac{\alpha  (-2+3 \alpha ) g'(\bar{x}_0) u''(\bar{x}_0)}{4 dt D g(\bar{x}_0){}^3 u'(\bar{x}_0)}
    \nonumber
  \\
  &
  \hspace*{-20mm}
    -
    \frac{3 (7-32 \alpha +32 \alpha ^2) u''(\bar{x}_0){}^2+4 (1-3 \alpha +3 \alpha ^2) u'(\bar{x}_0) u^{(3)}(\bar{x}_0)}{48 dt D g(\bar{x}_0){}^2 u'(\bar{x}_0){}^2}
    \bigg] \Delta x^4
    \nonumber
  \\
  &
  \hspace*{-30mm}
    +
    \frac{(1-2 \alpha )^2 u''(\bar{x}_0){}^2}{32 dt^2 D^2 g(\bar{x}_0){}^4 u'(\bar{x}_0){}^2}    
    \Delta x^6
    %
%
%
\; .
    \label{eq:devPxfromPu_tedious}
\end{align}%
\end{small}%
This form seems to be different from~(\ref{eq:resinfinitesimalpropag_nlin_dev}) because it still involves the function $u(x)$ that should not be present in the microscopic propagator for $x$ (unless of course the transformation is the identity $u(x)=x$ in which case~(\ref{eq:devPxfromPu_tedious}) is equal to~(\ref{eq:resinfinitesimalpropag_nlin_dev})).
However,  as checked with a direct computation, using the generalised substitution rules~(\ref{eq:Ito3})-(\ref{eq:Ito6}) allows one to remove 
all dependencies of~(\ref{eq:devPxfromPu_tedious}) in the function $u(x)$. Strikingly, the result is the correct propagator~(\ref{eq:resinfinitesimalpropag_nlin_dev}).

This computation shows that one can follow without inconsistencies the different branches of Fig.~\ref{fig:commutative-diag_non-lin-variable-change} for non-linear transformations, provided that the correct expansion is done when performing the change of variables in the action (yielding~(\ref{eq:devPxfromPu_tedious})) and that the  generalised substitution rules~(\ref{eq:Ito3})-(\ref{eq:Ito6}) are applied to the prefactor of the Gaussian weight
$
{|g(\bar x_t)|}^{-1}
\!
\exp
\!\big\{\!
\!-\!
\Delta x^2/[4D\,dt\,g(\bar x_t)]^2
\big\}
$%
, after the expansion of the infinitesimal propagator.

\subsection{Discussion}
\label{sec:disc-general-it=o-prescr}

In this subsection, we gather the previous results on the change of discretisation and 
the change of variables in a common description, aiming at understanding which are the valid rules of stochastic calculus that apply in the action.
We first describe the origin of the observed issues  
in the infinitesimal propagator, setting down modified substitution rules than can be applied 
``{inside}'' the exponential of the propagator (instead of ``outside'', on the prefactor of the Gaussian weight as done so far).
We then formulate a modified chain rule in continuous time that one
should apply in the path integral formalism.

\subsubsection{(Generalised) substitution rules and exponentials of infinitesimals}
\label{sec:gener-it=o-prescr}

In subsec.~\ref{sec:from-discr-anoth} and~\ref{sec:non-line-transf}, we noted that the expansion of the infinitesimal propagator involves a separation between 
\begin{enumerate}
\item
a purely Gaussian weight (which defines the probability distribution of the increment $\Delta x=x_{dt}-x_0$) and 
\item
a prefactor gathering all other terms, of the form $1+O(dt^{1/2})+O(dt)$.
\end{enumerate}
We now first show explicitly that the generalised substitution 
rules~(\ref{eq:Ito2b})-(\ref{eq:Ito6})  cannot be applied in the 
exponential and we elucidate which are the ``modified substitution 
rules'' to use in the exponentiated expression.
Recalling the notation,
$
 \mathcal N= 1/({4\pi D dt})^{1/2}
$
we denote by
\begin{align}
\PP^\GG_t
\
  \stackrel{\alpha}{=}
\
  \frac
  {\mathcal N}
  {|g(\bar x_t)|}
  \,\ee^{
  -\frac 12 \frac{dt}{2D} 
  \!
  \textnormal{
  $\big(\frac{\Delta x}{dt}\big)^2$
  }
  \!\!\!\!\big/ {g(\bar x_t)^2}
}
\end{align}
the part of the infinitesimal propagator (taken in a given $\alpha$-discretisation) which corresponds to the Gaussian distribution of $\Delta x$.
Then, either for the change of discretisation~(\ref{eq:resdev}) or for the non-linear change of variables~(\ref{eq:devPxfromPu_tedious}), the microscopic propagator is decomposed as
\begin{align}
\fl\qquad\
\mathbb P(x_{dt}|x_0)
\stackrel\alpha=
  \PP^\GG_0\!\times\!
  \exp
  \bigg\{
  A_0dt+A_1\Delta x + A_2 \Delta x^2 + A_3 \frac{\Delta x^3}{dt} + A_4 \frac{\Delta x^4}{dt}
  \bigg\}
\label{eq:decompgenericpropagator}
\end{align}
where $A_0,\ldots,A_4$ are functions of $\bar x_{0}$ taken in $\alpha$-discretisation.
The number of terms inside the exponential is finite, because higher-order powers of $\Delta x$ and $dt$ do not contribute at the orders $O(dt^{1/2})$ and $O(dt)$ we are interested in%
\footnote{%
Of course other computations than the change of discretisation and the change of variables that we considered in subsec.~\ref{sec:from-discr-anoth} 
and~\ref{sec:non-line-transf} could generate larger powers of $\Delta x$, such as $\Delta x^5/dt^2$ or $\Delta x^6/dt^2$, which are respectively of order $O(dt^{1/2})$ 
and $O(dt)$.
The modified rule that we present in the present subsection are easily adapted to such terms.
}%
.
Note that~(\ref{eq:resdev}) and~(\ref{eq:devPxfromPu_tedious}) are written in an expanded form, which goes up to order $\Delta x^6$ as
\begin{align}
\fl\quad\
  \exp
  \bigg\{
  A_0dt+&\!A_1\Delta x +A_2 \Delta x^2 + A_3 \frac{\Delta x^3}{dt} + A_4 \frac{\Delta x^4}{dt}
  \bigg\}
\nonumber
\\
&\hspace*{-10mm}
=
  1+
  A_0dt 
  +
  A_1 \Delta x
  +
  \Big(\frac 12{A_1^2}+A_2\Big)
  \Delta x ^2 
\nonumber
\\
& 
\qquad
 +
   A_3\frac{\Delta x ^3}{dt}
  +
  \big(A_1 A_3+A_4\big)\frac{\Delta x ^4 }{dt}
  +
  \frac{ A_3^2}{2}\frac{\Delta x ^6}{dt^2}
\;.
\end{align}
In this form, one can then apply the generalised substitution rules~(\ref{eq:Ito2b})-(\ref{eq:Ito6}) in a valid manner and 
reexponentiate the result, taking into account the orders in $dt$ correctly. (This is similar to what we have done 
in~(\ref{eq:devexpWN}) when treating the exponential of functions of the noise $\eta_t$ only that now we deal with a 
function of $\Delta x$.)
Denoting by $\sigma=2 D g^2(x)$ the noise amplitude,
one finds that the form~(\ref{eq:decompgenericpropagator}) of the propagator becomes
\begin{align}
\fl\quad
\mathbb P(x_{dt}|x_0)
\stackrel\alpha=
  \PP^\GG_0\!\times\!
  \exp
  \bigg\{
\!
  \Big[A_1+3A_3 \sigma\Big]\Delta x
+
  \Big[A_0+ A_2 \sigma + 3 (A_3)^2\sigma^2+3A_4 \sigma^2 \Big]dt
  \bigg\}
\label{eq:correct-propagator-dtDx}
\end{align}
with terms in the exponential that are order $\Delta x$ (or $dt^{1/2}$) 
and $dt$ only, as they should.

\smallskip

One observes by direct inspection that the generalised substitution rules~(\ref{eq:Ito2b})-(\ref{eq:Ito6}) 
\emph{cannot} be used directly inside the exponential of~(\ref{eq:decompgenericpropagator}) 
in order to get the correct result~\eqref{eq:correct-propagator-dtDx}.
Indeed, the term $A_3\frac{\Delta x ^3}{dt}$ in~(\ref{eq:decompgenericpropagator}) generates a 
quadratic contribution $\propto (A_3)^2$ in~\eqref{eq:correct-propagator-dtDx}.
The valid ``modified substitution rule'' to use in the exponential~(\ref{eq:decompgenericpropagator}) are thus
\\
\vspace*{-3mm}
\begin{align}
\fl\quad
\left.
\begin{minipage}{.85\linewidth}
\vspace*{-3mm}
\begin{align}
  \label{eq:Ito2MODIF}
\fl
  A_2(x) \Delta x^2 
  &
    \:\mapsto\:  A_2(x) \ 2Dg(x)^2 \,dt
  \\
\fl
  \label{eq:Ito3MODIF}
  A_3(x) \Delta x^3 \,dt^{-1} 
  &
    \:\mapsto\: 
                                3 A_3(x)  \ 2D g(x)^2  \Delta x+
                                3 \big[A_3(x)  \ 2D g(x)^2  \Delta x\big]^2 dt
  \\
\fl
  \label{eq:Ito4MODIF}
  A_4(x) \Delta x^4 \,dt^{-1} 
  &
    \:\mapsto\: 3 A_4(x) \  \big(2D g(x)^2\big)^2 \,dt
%
\end{align}
\end{minipage}
\right\}
\begin{minipage}{0.1\linewidth}
valid\\
only in\\
the exp.
\end{minipage}
\nonumber
\end{align}
\\
One observes that while the first and third line coincide with the corresponding ones in~(\ref{eq:Ito2b}) and (\ref{eq:Ito4}), 
the second line is different:
in~(\ref{eq:Ito3}) $\Delta x^3 dt^{-1}$ is substituted by an expression which is independent of its possible prefactor,
while in the exponential~(\ref{eq:decompgenericpropagator}) we need to use~(\ref{eq:Ito3MODIF}) that  \emph{effectively} replaces
$\Delta x^3 dt^{-1}$ by an expression which explicitly depends on its prefactor $A_3$ 
(in other words, the second term depends on $[A_3(x)]^2$).

In the formulation leading from~(\ref{eq:decompgenericpropagator}) to~(\ref{eq:correct-propagator-dtDx}) it is rather evident that the generalised substitution rules~(\ref{eq:Ito2b})-(\ref{eq:Ito6}) cannot be applied inside the exponential: indeed one can see $\ee^{A_3 \Delta x^3/dt}$  as equivalent to a moment-generating function of parameter $A_3$, and the exponent in~(\ref{eq:correct-propagator-dtDx}) as the corresponding cumulant-generating function, cut after $O(dt)$%
; thus, forgetting the quadratic term $\propto (A_3)^2$ in~(\ref{eq:Ito3MODIF}), which is of order $dt$, amounts to forgetting the term of degree~2 in the expansion of a cumulant-generating function%
\footnote{%
For $\Delta x^2$  and $\Delta x^4$, the higher-order term of the cumulant expansion do not contribute because they are $o(dt)$.
However, if a term in $A_5 \Delta x^5dt^{-2}$ had been present in~(\ref{eq:decompgenericpropagator}), its modified substitution rule in the exponential  would present a quadratic contribution $\propto (A_5)^2$ as in~(\ref{eq:Ito3MODIF}) for $A_3\Delta x^3dt^{-1}$.
}%
.

We note that Gervais and Jevicki~\cite{gervais_point_1976} have also determined in a quantum-field theory context that the correct procedure to change variables (in their case, to perform a canonical transformation) requires an expansion of the exponent up to terms of order $\Delta x^4 dt^{-1}$, akin to~(\ref{eq:decompgenericpropagator}). However, to our understanding, their treatment of these terms is unrelated to ours and remains perturbative in $D$, in contrast to our treatment which is non-perturbative.

\subsubsection{Modified chain rule}
\label{sec:modified-chain-rule}

The chain rule~(\ref{eq:chainrule_example}) allows one to deduce an $\alpha$-discretised Langevin equation on a variable $U(t)=u(x(t))$ from the corresponding Langevin equation on the variable $x(t)$ after a non-linear transformation, as discussed in Sec.~\ref{sec:non-line-transf}.
This same chain rule does not directly allow one to perform such non-linear change of variables at the level of the action 
(see App.~\ref{sec:inconsistency_action_chain-rule}).
To understand this issue on a general footing, let us start from a Langevin equation of the form~(\ref{eq:langevin_alpha_uU})
\begin{align}
  \label{eq:langevin_alpha_uUU}
  \dt U(t)
  \stackrel\alpha=
  &
  \
    F(U(t))+
    G(U(t))
    \eta(t)
\end{align}
The corresponding Onsager--Machlup weight reads
\begin{small}
\begin{align}
\fl
\quad
  \prod_t
\frac
{\mathcal N}
{|G(\bar U)|}
\times
\exp
\bigg\{
&
\!
-
\!
\int_0^{\tf}
\!\!
 dt
\bigg[
  \frac{1}{4D} 
  \bigg(
  \frac
  {\dt {U}-F (\bar U)+2\alpha D \,G(\bar U)G'(\bar U)}
  {G(\bar U)}
  \bigg)^2
  +    
  \alpha dt  F'(\bar U)
\bigg]
\bigg\}
\label{eq:OM_U_alpha}
\end{align}
\end{small}%
with each $\bar U=U(t)$ taken in $\alpha$-discretisation.
The naive approach consists in substituting $U(t)$ by $u(x(t))$ and then 
using the chain rule to determine the Onsager--Machlup weight for the trajectory $x(t)$.
In the next paragraphs, in order to understand why this procedure fails, we come back to the microscopic propagator~(\ref{eq:resinfinitesimalpropagx0bU}) corresponding to~(\ref{eq:OM_U_alpha}), in which we 
expand the square and we study separately the terms affine in~$\dt U$ and the term quadratic in~$\dt U$.
As we now show, the result is that the standard chain rule allows one to transform the terms affine in $\dt U$, while
to correctly transform the quadratic term $\propto (\dt U)^2$, one has to use a ``modified chain rule''.

\paragraphitalics{a Terms affine in $\dt U$}
\label{sec:terms-affine-part}
For the infinitesimal propagator, these terms  take the form
\begin{equation}
  \label{eq:affindtU}
  \mathcal B_1 = dt\,B_0(\bar U) + dt\, B_1(\bar U)\,\dt U
\end{equation}
where the first and second terms are of orders $O(dt)$ and $O(dt^{1/2})$, respectively. 
In order to reexpress $\ee^{\mathcal B_1}$ in terms of the original variable $x(t)$ one can follow either of the two following approaches.
\begin{enumerate}
\item 
In discrete time, one takes the same path as in the previous subsection.
Firstly, one discretises time explicitly; secondly, one expands $U_{dt}-U_0=U(x_{dt})-U(x_0)$ in powers of 
$\Delta x=x_{dt}-x_0$ around $\bar x_0$ using~(\ref{eq:X0})-(\ref{eq:Xdt}).
With the usual substitution rule~(\ref{eq:itopresc}) $\Delta x^2 = 2D g(x)^2 \,dt$,  after reexponentiation one obtains that
\begin{align}
\fl
  \ee^{\mathcal B_1}
  \stackrel\alpha
  =
  \exp
  \Big\{
&
    dt\,B_0\big(u(x(t)\big)
\nonumber
\\
\fl
&
    +
    dt\,B_1\big(u(x(t)\big)\Big[u'(x(t))\, \dt x(t)  +     (1-2\alpha)D\,g\big(x(t)\big)^2 u''\big(x(t)\big) \Big]
    \Big\}
  \label{eq:expA1chainr}
\end{align}
in the $dt\to 0$ limit.
In the light of subsec.~\ref{sec:direct-change-discr}, the computation involves no term in $\Delta x^3 dt^{-1}$ (nor higher order in 
powers of $\Delta x^n dt^m$), 
implying that the standard substitution rules could have also  been applied \emph{inside} the exponential.
\item In continuous time, one can use the chain rule~(\ref{eq:chainrule_example}) inside the exponential, for $U(t)=u(x(t))$ to get the result~\eqref{eq:expA1chainr}.
It is valid here as shown by the discrete-time computation described in the previous point.
\end{enumerate}

\paragraphitalics{b Term proportional to $({\rm d}_t U)^2$}
\label{sec:terms-prop-part}
This term takes the form
\begin{equation}
  \label{eq:quaddtU}
  \mathcal B_2 = - \frac{1}{2} \frac {dt}{2D} \, B_2(\bar U)\, \big({\rm d}_t U\big)^2
  \qquad
  \text{with }\
  B_2= \frac{1}{G^2}
  \;.
\end{equation}
It is of \emph{order $dt^0$} and if one naively uses the chain rule to compute ${\rm d}_t U = {\rm d}_t [u(x(t))]$, 
one misses a number of terms;
such computation would yield
\begin{align}
  \label{eq:wrongexpquad}
\fl
  \ee^{\mathcal B_2}
  \!
  \stackrel{{\text{\bf{wrong}!}}} =
  \!
  \exp
  \!
  \bigg\{
  \!
  -\frac {dt}{4D}B_2\big(u(x(t)\big) \Big[u'(x(t)) {\rm d}_t x(t)  +     (1-2\alpha)D\,g\big(x(t)\big)^2 u''\big(x(t)\big) \Big]^2
  \bigg\}
\end{align}
where $g(x)^2=1/B_2(u(x))$.
Instead, one should discretise in time, using
\begin{equation}
  \label{eq:A1U0}
  \mathcal B_2 = -\frac{1}{4D} dt^{-1}\, B_2(\bar U_0)\, \big(U_{dt}-U_0)^2
  \qquad\textnormal{as~~$dt\to 0$}
\end{equation}
where 
$U_{0}=u(x_{0}),\
U_{dt}=u(x_{dt})$ and $\bar U_0$ is defined in~(\ref{eq:Ubar0}).
We also define the function  
  $b_2(x)=B_2(u(x))=1/g(x)^2$
for lighter notations.

\bigskip
\emph{Expansion of $\mathcal B_2${~--~}}
Using the relations~(\ref{eq:Ubar0})-(\ref{eq:Xdt}), one then expands~(\ref{eq:A1U0}) in powers of $\Delta x$ up to order $dt$ to find
\begin{small}
\begin{align}
\fl\quad\qquad
  \mathcal B_2=&
-\frac{1}{4 D g(\bar{x}_0){}^2} \: \frac{\Delta x ^2}{dt}
\ +\
\frac{(-1+2 \alpha )u''(\bar{x}_0)}{4  D g(\bar{x}_0){}^2 u'(\bar{x}_0)} \:\frac{\Delta x ^3 }{dt}
\nonumber
\\
\fl
&
+
\bigg[-\frac{(-1+\alpha ) \alpha  g'(\bar{x}_0) u''(\bar{x}_0)}{4  D g(\bar{x}_0){}^3 u'(\bar{x}_0)}-\frac{(1+8 (-1+\alpha ) \alpha ) u''(\bar{x}_0){}^2}{16  D g(\bar{x}_0){}^2 u'(\bar{x}_0){}^2}
\nonumber
\\
\fl
&
\hspace*{4.34cm}
+\frac{(-1-3 (-1+\alpha ) \alpha ) u^{(3)}(\bar{x}_0)}{12  D g(\bar{x}_0){}^2 u'(\bar{x}_0)}
\bigg]
\,\frac{\Delta x ^4 }{dt}
  \label{eq:A2dev}
\;,
\end{align}
\end{small}%
which is not obviously related to~(\ref{eq:wrongexpquad}).
We note that this expression contains a crucial term proportional to $\Delta x ^3dt^{-1}$ which, as we have discussed in subsec.~\ref{sec:gener-it=o-prescr}, has to be treated with great care. The modified substitution rule~(\ref{eq:Ito3MODIF})  has to be used here [and not the rule~(\ref{eq:Ito3})], in order to handle correctly $\Delta x ^3dt^{-1}$ \emph{inside} the exponential.
%
We also remark that the term in $\Delta x ^3dt^{-1}$ \emph{is non-zero for an additive noise} (\emph{i.e.}~when $g(x)$ is constant), indicating that non-linear changes of variables also have to be handled with care in this case.

\bigskip

\emph{Expansion of $\ee^{\mathcal B_2}${~--~}}
The correct procedure to follow in order to first use the (simple) substitution rules~(\ref{eq:Ito2b})-(\ref{eq:Ito6}) for $\Delta x^n$ is to first expand the terms of~(\ref{eq:A2dev}) which are not in $\Delta x^2dt^{-1}$, and to use then the substitution rules~(\ref{eq:Ito2b})-(\ref{eq:Ito6}). Alternatively, one can use the modified ones~(\ref{eq:Ito2MODIF})-(\ref{eq:Ito4MODIF}) which are valid inside an exponential.
Recalling the notation $g^2=1/b_2$, and writing
\begin{align}
\ee^{(\ldots)\Delta x^2}=\exp\Big[-\frac 12 \frac{b_2(\bar x_0)\,dt}{2D}\Big(\frac{\Delta x}{dt}\Big)^2 \big(u'(\bar x_0)\big)^2\Big]
\end{align}
one obtains
  \begin{align}
\fl
    \frac{\ee^{\mathcal B_2}}{\ee^{(\ldots)\Delta x^2}}
    = \exp
    \bigg\{
&
\frac{3}{2} (-1+2 \alpha )   u'(\bar{x}_0) u''(\bar{x}_0) \Delta x
\nonumber
\\
&
\!\!\!\!
+
  dt
  \bigg[
  -\frac{3 D (1-2 \alpha )^2 u''(\bar{x}_0){}^2}{4 b_2(\bar{x}_0)}+\frac{3 D (1-2 \alpha )^2 u'(\bar{x}_0){}^2 u''(\bar{x}_0){}^2}{2 b_2(\bar{x}_0)}
  \label{eq:reexpA2devC}
\\
&
\quad
\quad
+
u'(\bar{x}_0)
\bigg(
\frac{3 D (-1+\alpha ) \alpha  b_2'(\bar{x}_0) u''(\bar{x}_0)}{2 b_2(\bar{x}_0){}^2}
\nonumber\\
&
\quad\quad
-\frac{D (1+3 (-1+\alpha ) \alpha ) u^{(3)}(\bar{x}_0)}{b_2(\bar{x}_0)}
\bigg)
  \bigg]
  \bigg\}
\;.
\nonumber
  \end{align}%
This result is completely different from the naive result~(\ref{eq:wrongexpquad}), obtained from the use of 
the chain rule~(\ref{eq:chainrule_example}) in the exponential, that can be recast as 
\begin{small}%
  \begin{align}
\fl
\qquad
    \frac{\ee^{\mathcal B_2}}{\ee^{(\ldots)\Delta x^2}}
  \!
  \stackrel{\text{\bf{wrong}!}} =
  \!
\exp
  \bigg\{
&
  \frac{1}{2} (-1+2 \alpha ) u'(\bar{x}_0) u''(\bar{x}_0) 
  \Delta x 
  -
  \frac{ D (1-2 \alpha )^2 u''(\bar{x}_0){}^2}{4 b_2(\bar{x}_0)}
  dt
  \bigg\}
\;.
\label{eq:reexpA2devCWRONG}
\end{align}%
\end{small}%
(They coincide for linear transformations such that $u''=0$.)

The result~(\ref{eq:reexpA2devC}) allows one to identify the correct (but complicated) 
form of the chain rule to be used in the exponential for terms of the form~(\ref{eq:quaddtU}).
%
%
Instead of the chain rule~(\ref{eq:chainrule_example}) that would lead to~(\ref{eq:reexpA2devCWRONG}), one has  that
\\
%
\begin{minipage}{1.0\linewidth}%
\begin{align}%
\fl
  - \frac 12 \frac {dt}{2D}
 \, B_2&(U)\,
    \big({\rm d}_t U\big)^2
\nonumber
\\
\fl
&
\hspace*{-9.8mm}
  \stackrel{\text{in\ exp.}}\mapsto
-\,\frac 12 \frac{dt}{2D}\,b_2(x)\,(\dt x )^2
  (u'(x))^2
\nonumber
\\
\fl
&
+
\frac{3}{2} (-1+2 \alpha )   u'(x ) u''(x ) dt\, \dt x
  \label{eq:reexpA2devC2}
\\
\fl
&
+
  dt
  \bigg[
  -\frac{3 D (1-2 \alpha )^2 u''(x ){}^2}{4 b_2(x )}+\frac{3 D (1-2 \alpha )^2 u'(x ){}^2 u''(x ){}^2}{2 b_2(x )}
\nonumber
\\
\fl
&
\quad
\quad\ \
+
u'(x )
\bigg(
\frac{3 D (-1+\alpha ) \alpha  b_2'(x ) u''(x )}{2 b_2(x ){}^2}-\frac{D (1+3 (-1+\alpha ) \alpha ) u^{(3)}(x )}{b_2(x )}
\bigg)
  \bigg]
\;.
\nonumber
\end{align}%
%
\end{minipage}%
\\
(We took the $dt\to 0$ limit, with the r.h.s. being $\alpha$-discretised.)
In order to apprehend 
better the difference with the naive application of the chain rule~(\ref{eq:chainrule_example}), one can rewrite this result as
\begin{align}
\fl
  - \frac 12 \frac {dt}{2D}
 \, B_2&(U)\,
    \big({\rm d}_t U\big)^2
\nonumber
\\[-5mm]
\fl
&
\hspace*{-9.8mm}
  \stackrel{\text{in\ exp.}}\mapsto
-\,\frac 12 \frac{dt}{2D}
  \,b_2(x)
  \Big[
\overbrace{
\vphantom{\Big|}
u'(x)\dt  x + (1-2\alpha)D\,g(x)^2 u''(x)
}^{\text{chain~rule~(\ref{eq:chainrule_example})}}
  \Big]^2
\nonumber
\\[1mm]
\fl
&
+
(-1+2 \alpha )  u'(x) u''(x)\,dt\,\dt x
  \label{eq:reexpA2devC2diff}
\\[1mm]
\fl
&
+
  dt
  \bigg[
\frac{3 D (-1+\alpha ) \alpha  u'(x) b_2'(x) u''(x)}{2 b_2(x){}^2}
\nonumber
\\
\fl
&
\quad
\quad\ \
+\frac{D ((1-2 \alpha )^2 (-1+3 u'(x){}^2) u''(x){}^2-2 (1+3 (-1+\alpha ) \alpha ) u'(x) u^{(3)}(x))}{2 b_2(x)}
  \bigg]
\;,
\nonumber
\end{align}%
the three last lines being the terms one misses if one merely applies~(\ref{eq:chainrule_example}).
\bigskip

\emph{Special cases{~--~}} 
One notes that this modified chain rule remains non-trivial in the three following simplified cases:
\begin{itemize}
\item Stratonovich discretisation ($\alpha=1/2$):
\begin{align}
\fl
\hspace{-3mm}
  - \frac 12 \frac {dt}{2D}
 \, B_2&(U)\,
    \big({\rm d}_t U\big)^2
\nonumber
\\
\fl
&
\hspace*{-11mm}
  \stackrel{\text{in\ exp.}}\mapsto
-\,\frac 12 \frac{dt}{2D}
  \,b_2(x)
  \big[
u'(x)\dt  x 
  \big]^2
-
  D\,dt
  \bigg[
\frac{3 u'(x) b_2'(x) u''(x)}{8 b_2(x){}^2}
+
\frac{u'(x) u^{(3)}(x)}{4 b_2(x)}
  \bigg]\:.
\label{eq:reexpA2devC2diffStrato}
\end{align}%

\item Additive noise ($B_2(U)=B_2=b_2(x)=b_2=1/g^2(x)=1/g^2$):
\begin{align}
\fl
\hspace{-3mm}
  - \frac 12 \frac {dt}{2D}
 \, b_2& \: 
    \big({\rm d}_t U\big)^2
\nonumber
\\
\fl
&
\hspace*{-11mm}
  \stackrel{\text{in\ exp.}}\mapsto
-\,\frac 12 \frac{dt}{2D}
  \,b_2 \:
  \Big[
u'(x)\dt  x 
+ (1-2\alpha)D\,g^2 u''(x)
  \Big]^2
\label{eq:reexpA2devC2diffadditnoise}
\\[1mm]
\fl
&
+
(-1+2 \alpha )  u'(x) u''(x)\,dt\,\dt x
%
\nonumber
\\[1mm]
\fl
&
+
  \frac{D\,dt}{2 b_2}
  \bigg[
{(1-2 \alpha )^2 (-1+3 u'(x){}^2) u''(x){}^2-2 (1+3 (-1+\alpha ) \alpha ) u'(x) u^{(3)}(x)}
  \bigg]
\;.
\nonumber
\end{align}%

\item Additive noise and Stratonovich discretisation:
\begin{align}
\fl
  - \frac 12 \frac {dt}{2D}
 \, b_2& \:
    \big({\rm d}_t U\big)^2
\
  \stackrel{\text{in\ exp.}}\mapsto
\
-\,\frac 12 \frac{dt}{2D}
  \,b_2 \:
  \big[
u'(x)\dt  x 
  \big]^2
-D\,dt\,\frac{ u'(x) u^{(3)}(x)}{4 b_2}
\;.
\label{eq:reexpA2devC2diffStratoaddit}
\end{align}%
This last case is peculiarly striking, because one could have expected  
the standard chain rule of differentiable calculus to be valid in the dynamical action of an 
additive-noise Stratonovich-discretised Langevin equation (as it is valid at the Langevin equation level). 
Surprisingly, this is not the case as soon as $u^{(3)}(x)\neq 0$.
\end{itemize}

\bigskip
\section{Outlook}
\label{sec:outlook}

The trajectory probability of Langevin processes is well described by a path-integral weight, through either the MSRJD~\cite{Janssen1976,BaJaWa76} or the 
Onsager--Machlup~\cite{onsager_fluctuations_1953,machlup_fluctuations_1953II} formulations.
In this article we studied the behaviour of the Langevin equation and its corresponding Onsager--Machlup action under two generic 
transformations: a change of $\alpha$-discretisation and a non-linear change of variables. 
The correct rules to perform these transformations at the level of  the Langevin equations are well-known,  they have been 
recalled in this article, and we verified, once again, that they are reversible.

Consistency requires that the trajectory probability constructed from the Langevin equation of a variable $u(t)=u(x(t))$ in a discretisation 
scheme $\bar\alpha$ be the same as the  trajectory probability of the Langevin process 
of the variable $x(t)$ in another discretisation scheme $\alpha$, after applying to the latter the corresponding discretisation and non-linear transformations.
Figures~\ref{fig:commutative-diag_change-of-discretisation} and~\ref{fig:commutative-diag_non-lin-variable-change}
provide sketches of this statement for the discretisation scheme transformation and the non-linear transformation, respectively. 
However, it was observed in the literature that their use in the action could yield inconsistencies, 
both in the stochastic field-theory context~\cite{Langouche81,aron_dynamical_2014_arxiv1} and in the quantum-mechanical one~\cite{gervais_point_1976,Sa77,LaRoTi79,Tirapegui82,AlDa90,ApOr96}. The aim of the present article was to identify the 
generalisation of the Itō rule and the correct rules of calculus that ensure the reversibility of the construction.

By carefully analysing the discrete-time behaviour of the propagator corresponding to the infinitesimal evolution during 
a time step $dt\to 0$, we identified the source of inconsistencies and we provided procedures that allow one to 
perform the transformations in the action in a correct manner. 

To summarise them, we now list the possible
sources of issues. At the infinitesimal level, we denote the trajectory increment by $\Delta x=x_{t+dt}-x_t$ which is typically of order $dt^{1/2}$.
The main source of problems is that terms of the form $\Delta x^3dt^{-1}$,  $\Delta x^4dt^{-1}$ and $\Delta x^6dt^{-2}$ are generated in the infinitesimal propagator upon the mentioned transformations, while they are not generated at the Langevin level.
First, they have to be correctly identified, and second, one has to understand their behaviour in the $dt\to 0 $ limit.
We have provided generalised substitution rules~(\ref{eq:Ito2b})-(\ref{eq:Ito6}) that allow one to do so (they generalise the usual Itō prescription 
$dB_t^2=dt$ for the Brownian motion).
An important point is that these relatively simple
rules have to be used in the prefactor of the Gaussian weight of the infinitesimal 
propagator (after a $dt\to 0$ expansion), and not inside the exponential of this propagator.
We have provided a simple explanation of this condition in subsec.~\ref{sec:gener-it=o-prescr}.
If one insisted upon applying the transformations in the exponential, 
the modified substitution rules become significantly more complicated and are given in Eqs.~(\ref{eq:Ito2MODIF})-(\ref{eq:Ito4MODIF}). 
%

In the continuous-time path integral, an important 
consequence of the previous observations is that one cannot use the stochastic chain rule~(\ref{eq:chainrule_example})  to perform changes of variables. 
One has, instead, to rely on a time-discrete expansion or on a modified chain rule, described in subsec.~\ref{sec:modified-chain-rule}.
We emphasise that the application of the invalid chain rule~(\ref{eq:chainrule_example}) in the action yields wrong results 
\emph{even for an additive-noise Stratonovich-discretised Langevin equation}. The reason for this is that under a non-linear transformation of variables the equation 
becomes one with multiplicative noise.

For future perspectives, we can list a number of interesting questions to address:
\begin{enumerate}

\item
It would be helpful to identify similar rules that would solve inconsistencies observed when manipulating the 
MSRJD action~\cite{aron_dynamical_2014_arxiv1}, because many field theories (including quantum ones) are better written in this formalism or in similar ones that also involve a response field.
%
%

%
\item
The generalisation to more than one degree of freedom could be tricky~\cite{lau_state-dependent_2007} but 
should be very interesting and useful.

\item
Langevin equations with inertia (a second time derivative) and/or coloured noise 
approach in the overdamped and/or white noise limit the 
equation that we studied here in the Stratonovich scheme
(see, \emph{e.g.}~\cite{gardiner_handbook_1994,Aron_etal_2010}).
It would be interesting to understand how the issues discussed in the present article
arise and are solved in these regularised and better behaved cases since, as we showed, 
even the action in the Stratonovich discretisation scheme has to be treated attentively.

\item
The results we have presented also encourage one to revisit the validity of some non-linear transformation 
used in quantum field theory~\cite{gervais_point_1976,Sa77,LaRoTi79,Tirapegui82,AlDa90,ApOr96}, where 
the Lagrangians defining the action take forms that are similar to that of statistical mechanics.

\end{enumerate}

\bigskip

\emph{Acknowledgements.}
We are deeply indebted to Maxence~Ernoult, in collaboration with whom we initiated this research work.
LFC gratefully thanks Camille Aron and Gustavo Lozano for early discussions on this problem. 
She is a member of Institut Universitaire de France.
VL gratefully thanks Eric Bertin for fruitful discussions on stochastic calculus, 
and acknowledges support by the ERC Starting Grant 680275 MALIG, by the ANR-15-CE40-0020-03 Grant LSD and by the UGA IRS PHEMIN project.
%


\appendix
\section*{Appendices}

\renewcommand{\thesection}{A}
\section{Determination of the infinitesimal propagator: other approaches}
\label{sec:infin-prop-other}

In this appendix, in order to shed a different light on the use of the Itō prescription 
in the determination of the infinitesimal propagator, we review other less pedestrian 
approaches than the one presented in Sec.~\ref{sec:infin-prop-path}.

\subsection{\emph{À la} Lau--Lubensky}
\label{sec:much-simpl-appr}

To compute $  \delta(x_{dt}-X_1(x_0,\eta_0)) $ in~(\ref{eq:Px1eta0}), it proves simpler~\cite{lau_state-dependent_2007} to start from the following identity, where the argument of the first delta is the equation of motion at $t=0$:
\begin{align}
\fl\qquad
\delta
\bigg(
\overbrace{
  \eta_0-
  \frac
  {\frac{x_{dt}-x_0}{dt}-f (\bar x_0 )}
  {g(\bar x_0)}
}^{
  \equiv F(\eta_0,x_0,x_{dt})
  }
\bigg)
&\stackrel{\eqref{eq:x1X1}}=
\frac1
{|\partial_{x_{dt}}F(\eta_0,x_0,x_{dt})|}
\delta\big(x_{dt}-X_1(x_0,\eta_0)\big)
\; , 
\end{align}
where one recognises $F(\eta_0,x_0,x_{dt})=\eta_0-H_0(x_0,x_{dt})$ from Eq.~(\ref{eq:solH0notexpanded}). One thus has
\begin{align}
\fl\qquad
|\partial_{x_{dt}}H_0(x_0,x_{dt})|\:
\delta\big(\eta_0-H_0(x_0,x_{dt})\big)
&\stackrel{\eqref{eq:solH0notexpanded}}=
\delta\big(x_{dt}-X_1(x_0,\eta_0)\big)
\; , 
\end{align}
so that finally
\begin{align}
\fl\qquad
  \mathbb P(x_{dt}|x_0) 
&\stackrel{\eqref{eq:Px1eta0}}= 
  \int d\eta_0\: 
  |\partial_{x_{dt}}H_0(x_0,x_{dt})|\:
  \delta\big(\eta_0-H_0(x_0,x_{dt})\big)\:
  P_\text{noise}(\eta_0) 
  \; . 
\label{eq:prob-trick}
\end{align}
By direct computation, one obtains
\begin{align}
  \label{eq:derH0}
\fl\qquad
  \partial_{x_{dt}}\! H_0(x_0,x_{dt})
  &=
    \frac 1{dt} 
    \frac 1{g(\bar x_0)}
    \Big[
    1-\alpha dt\,f'(\bar x_0)-\big(x_{dt}-x_0-f(\bar x_0)\,dt\big)\alpha\tfrac{g'(\bar x_0)}{g(\bar x_0)}
    \Big]
\end{align}
that, using the Dirac delta in~\eqref{eq:prob-trick} to re-express $x_{dt}-x_0$ as a function of~$\eta_0$,
implies 
\begin{align}
\fl\qquad
  |\partial_{x_{dt}}H_0(x_0,&x_{dt})|\:
  \delta\big(\eta_0-H_0(x_0,x_{dt})\big)\:
\nonumber
\\
&
\stackrel{(\ref{eq:H0})}
=
  \frac 1{dt} 
  \frac 1{|g(\bar x_0)|}
    \Big[
    1-\alpha dt\,f'(\bar x_0)-\eta_0 g'(\bar x_0)\alpha dt 
    \Big]
\\
&\stackrel{(\ref{eq:devexpWN})}=
  \frac 1{dt} 
  \frac 1{|g(\bar x_0)|}
  \ee^{-\alpha dt\,f'(\bar x_0)-\eta_0 g'(\bar x_0)\alpha dt
  -D [g'(\bar x_0)]^2 \alpha^2 dt}
  \; . 
\label{eq:JacLauLub}
\end{align}
Inserting this expression in Eq.~(\ref{eq:prob-trick}), one finds exactly 
the same propagator given in Eq.~(\ref{eq:resinfinitesimalpropagx0b}).
This provides a justification for the use of the Itō rule~(\ref{eq:itopresc_eta}) 
in Eqs.~(\ref{eq:invdhoX1b}), (\ref{eq:invdhoX1d}) and~(\ref{eq:gx0gx0barexp}), used in the derivation 
of the propagator presented in Sec.~\ref{sec:infin-prop-path}.

Last, we mention that Lau and Lubensky~\cite{lau_state-dependent_2007} actually follow a slightly different route, which involves a Fourier transformation, but in the end their treatment is equivalent to the one we presented in this paragraph.

\subsection{\emph{\`A la} Itami--Sasa}
\label{sec:anoth-comp-jacob-IS}

In order to calculate the Jacobian $\frac{1}{|\partial_{\eta_{0}}X_{1}(x_{0},H_{0})|}$ arising in~(\ref{eq:Px1x0Pnoise}), one can proceed as follows~\cite{itami_universal_2017}: we write the first time step $0 \curvearrowright dt$ of the equation of motion as
\begin{align}
X_{1}(x_{0},\eta_{0})=x_{0}&+f[
\overbrace{\alpha X_{1}(x_{0},\eta_{0})+(1-\alpha)x_{0}}^{\bar x_0}
]dt
\nonumber
\\
&
+g[\alpha X_{1}(x_{0},\eta_{0})+(1-\alpha)x_{0}]\eta_{0}dt
\; . 
\end{align}
Differentiating with respect to the noise, one obtains
\begin{align}
\partial_{\eta_{0}}X_{1}=\alpha\partial_{\eta_{0}}X_{1}f'( \bar x_0)dt
+\alpha\partial_{\eta_{0}}X_{1}g'( \bar x_0)\eta_{0}dt+g( \bar x_0)dt
\end{align}
that implies 
\begin{align}
\frac{1}{|\partial_{\eta_{0}}X_{1}|}&=\frac{1}{|g( \bar x_0)|dt}(1-\alpha f'( \bar x_0)dt-\alpha g'( \bar x_0)\eta_{0}dt)
\; . 
\end{align}
Note that so far, no expansion nor approximation has been done: this result is exact. 
In order to exponentiate the numerator of this expression, one uses~(\ref{eq:devexpWN}):
\begin{align}
\frac{1}{|\partial_{\eta_{0}}X_{1}|}
&=\frac{1}{|g( \bar x_0)|dt}\exp[-\alpha f'( \bar x_0)dt-\alpha g'( \bar x_0)\eta_{0}dt -D\alpha^{2}g'( \bar x_0)^{2}dt]
\;.
\nonumber
\end{align}
This is the same expression as the one in Eq.~\eqref{eq:JacLauLub} 
obtained following the Lau--Lubensky approach, and the one that we obtained in 
Sec.~\ref{sec:infin-prop-path}.

\subsection{A continuous-time derivation of the Jacobian}
\label{sec:anoth-comp-jacob-Arnold}

In the quantum-mechanical context a continuous-time formalism is used
and the subtleties linked to the discretisation scheme are 
usually encoded in the choice of the value of the Heaviside theta function at zero, 
$\Theta(0)=\alpha$~\cite{zinn-justin_quantum_2002}.
In this field, the Jacobian $|\partial_{\eta_0} X_1(x_0, H_0(x_0,x_{dt}))|$
is computed with the help of the identity $\det(1+ C_{\eta_0}) = \exp \Tr \ln(1+C_{\eta_0})$
where $C_{\eta_0}$ is the part of the Jacobian that depends on the noise. The  
expression $\ln(1+C_{\eta_0})$ is further expanded in powers of $C_{\eta_0}$ to quadratic order (so as to keep terms that are 
quadratic in the noise and contribute to the trace involving a time integral when the noise is delta correlated)~\cite{Ar00}. 
The explicit calculation of the Jacobian along these lines was explained in App.~D in~\cite{aron_dynamical_2016} and 
constitutes another way of arriving at the expression in Eq.~\eqref{eq:JacLauLub}. It is less useful for our purposes in this 
article since it works in continuous time and does not allow to make immediate contact with the (generalised) substitution rules in 
discrete time.

\renewcommand{\thesection}{B}
\section{Justifying the generalised substitution rules}
\label{sec:just-gener-it=o}

\subsection{The usual $\Delta x^2 = 2D g(x)^2 \,dt$ substitution}
\label{sec:just-it=o}
Stochastic calculus tells us that, when expanding infinitesimals, for a standard Brownian motion $B_t$, one has:
\begin{align}
  \label{eq:SCalculus}
  dB_t^2=dt
  \; . 
\end{align}
For our time-discrete noise, $\eta^2_t =2D/dt$. For a more complex variable such as $x$, the solution 
of the Langevin equation~(\ref{eq:langevin_alpha}), the substitution rule~(\ref{eq:itopresc}) implies
\begin{align}
  \label{eq:Ito2}
  \Delta x^2 = 2D g(x)^2 dt  \quad \textnormal{($+O(dt^{3/2})$ as $dt\to 0$)}
\end{align}
where on the r.h.s., the argument $x$ of $g(x)$ can be taken at any discretisation point, at minimal order in $dt$.
As discussed in Sec.~\ref{sec:chain-rule}, there is no direct argument on the distribution of $\Delta x$
which allows one to use~(\ref{eq:Ito2}) point-wise.
The meaning of this relation is to be found in an integral way.
Following Øksendal~\cite{oksendal_stochastic_2013}, one uses the following ingredients:
\begin{itemize}

\item Two functions $A_1$ and $A_2$ of the process $x$ are equivalent if the $L^2$ norm of the temporal integral of their difference is zero:
  \begin{align}
\fl
    A_1[\dt  x(t),x(t)]=&A_2[\dt  x(t),x(t)]
\nonumber
\\
&
\hspace*{-8mm}
      \ \Leftrightarrow \
      \bigg\langle
      \Big(
      \int_0^{\tf} dt\: \big\{ A_1[\dt  x(t),x(t)] - A_2[\dt  x(t),x(t)] \big\}
      \Big)^2
      \bigg\rangle
      \ = \ 0
      \label{eq:eqprocessint}
    \\
    &
\hspace*{-8mm}
      \ \Leftrightarrow \
      \bigg\langle
      \Big( \sum_t dt\:\big\{ A_1\big [\tfrac{\Delta x}{dt},x_t\big] - A_2\big [\tfrac{\Delta x}{dt},x_t\big] \big\} \Big)^2
      \bigg\rangle
      \stackrel{dt\to 0}\longrightarrow\: 0
      \label{eq:eqprocesssum}
      \;.
  \end{align}

\item Two Brownian increments $B_{t+dt}-B_t$ and $B_{t'+dt}-B_{t'}$  at different times $t\neq t'$ are independent:
  \begin{align}
    \label{eq:indepBt}
\fl
    \big\langle (B_{t+dt}-B_t)(B_{t'+dt}-B_{t'})\big\rangle
    = 
    \big\langle (B_{t+dt}-B_t)\big\rangle\big\langle(B_{t'+dt}-B_{t'})\big\rangle
    \quad
    \textnormal{if $t\neq t'$}
      \;.
  \end{align}
  
\item The following averages are computed (\emph{e.g.}~à la Wick) using the Gaussian nature of $B_t$:
  \begin{align}
    \label{eq:vmB2}
\fl
    \big\langle (B_{t+dt}-B_t)^2\big\rangle &= dt
    \; , 
    \\
    \label{eq:vmB4}
\fl
    \big\langle (B_{t+dt}-B_t)^4\big\rangle &= 3\, dt^2
    \; . 
  \end{align}
\end{itemize}

\bigskip
Let us thus show that in the sense of~(\ref{eq:eqprocessint})-(\ref{eq:eqprocesssum}), one has  $\Delta x^2 = 2D g(x)^2 \,dt$. For this, one computes
\begin{align}
\fl\quad  
  \bigg\langle
  \Big(
  \sum_t
  \big\{
   \Delta x^2 - &2D g(x_t)^2 \,dt
  \big\}
  \Big)^2
  \bigg\rangle
  \nonumber
\\
&  \stackrel{(\ref{eq:langevin_alpha})}
  =
  \bigg\langle
  \Big(
  \sum_t
  \big\{
   \big(dt f(x_t)+g(x_t)\eta_tdt\big)^2 - 2D g(x_t)^2 \,dt
  \big\}
  \Big)^2
  \bigg\rangle
    \nonumber
    \\
&  \stackrel{\phantom{(\ref{eq:langevin_alpha})}}
  =
  \bigg\langle
  \Big(
  \sum_t
  \big\{
   \big(g(x_t)\eta_tdt\big)^2 - 2D g(x_t)^2 \,dt
  \big\}
  \Big)^2
  \bigg\rangle
  +O(dt)
  \label{eq:Oks_1}
  \\
&  \stackrel{\phantom{(\ref{eq:langevin_alpha})}}
  =
  \bigg\langle
  \Big(
  \sum_t
   \big[(B_{t+dt}-B_t\big)^2 - dt\big] 2D g(x_t)^2 
  \Big)^2
  \bigg\rangle
  +O(dt)
  \label{eq:Oks_2}
  \\
&  \stackrel{\text{(\ref{eq:indepBt})}}
  =
  \sum_t
  \bigg\langle
  \Big(
   \big[(B_{t+dt}-B_t\big)^2 - dt\big] 2D g(x_t)^2 
  \Big)^2
  \bigg\rangle
  +O(dt)
  \nonumber
  \\
&
  \qquad
  +
  \sum_{t\neq t'}
  \bigg\langle
   \big[(B_{t+dt}-B_t\big)^2 - dt\big] 2D g(x_t)^2
  \bigg\rangle
  \label{eq:indepdB2}
  \\
&
  \ \qquad
  \phantom{
  +
  \sum_{t\neq t'}
  }
  \times
  \bigg\langle
   \big[(B_{t'+dt}-B_{t'}\big)^2 - dt\big] 2D g(x_{t'})^2
  \bigg\rangle
  \nonumber
  \\
&  \stackrel{\text{(\ref{eq:vmB2})}}
  =
  \sum_t
  \underbrace{
  \bigg\langle
  \Big[(B_{t+dt}-B_t\big)^2 - dt\Big]^2\bigg\rangle
  }_{
  {\displaystyle\stackrel{\textnormal{\tiny{(\ref{eq:vmB2})-(\ref{eq:vmB4})}}} = }
  3dt^2-2dt^2+dt^2
  }
  \Big\langle \big(2D g(x_t)^2\big)^2
  \Big\rangle
  +O(dt)
  \label{eq:afterindepdB2}
  \\
&  \stackrel{\phantom{(\ref{eq:vmB2})}}
  =
  dt \sum_t 2 dt
  \Big\langle \big(2D g(x_t)^2\big)^2
  \Big\rangle
  +O(dt)
    \\
&  \stackrel{\phantom{(\ref{eq:vmB2})}}
  =
  O(dt)
  \label{eq:finaldt_O2}
\end{align}
which goes to zero as $dt\to 0$, hence finishing the proof of~(\ref{eq:Ito2}).

\medskip

Note that when going from~\eqref{eq:indepdB2} to~\eqref{eq:afterindepdB2}, 
one cancels the sum over different time indices $t\neq t'$ using that $x_t$ is independent of $B_{t+dt}-B_t$:
\begin{align}
  \label{eq:factorBrown}
\fl\quad
    \bigg\langle
   \big[(B_{t+dt}-B_t\big)^2 - dt\big] 2D g(x_t)^2 
  \bigg\rangle
  &
    \stackrel{\phantom{(\ref{eq:vmB2})}}=
    \bigg\langle
    (B_{t+dt}-B_t\big)^2 - dt
    \bigg\rangle
    \bigg\langle
     2D g(x_t)^2 
  \bigg\rangle
\nonumber
  \\
  &
    \stackrel{\text{(\ref{eq:vmB2})}}=
    0
    \; . 
    \label{eq:zero_average}
\end{align}
In particular, the factor $2$ in $\Delta x^2 = 2D g(x)^2 \,dt$ is essential, because it allows one to factorise by $2D g(x)^2$ between~(\ref{eq:Oks_1}) and~(\ref{eq:Oks_2}), and to obtain \emph{in fine}  the cancellation in~\eqref{eq:zero_average} which makes that~\eqref{eq:finaldt_O2} is of order $dt$.

\subsection{The generalised substitution rule $\Delta x^4 dt^{-1} = 3\: \left(2D g(x)^2\right)^2 \,dt$}
One follows the same path, using
$
    \big\langle (B_{t+dt}-B_t)^8\big\rangle = 105\, dt^4\,,
$
one computes
\begin{small}
\begin{align}
\fl\quad 
  \bigg\langle
  \Big(
  \sum_t
  \Big\{
   \tfrac{\Delta x^4}{dt} - & 3\,\big(2D  g(x_t)^2\big)^2 \,dt
  \Big\}
  \Big)^2
  \bigg\rangle
\nonumber
  \\
  &
  \stackrel{(\ref{eq:langevin_alpha})}
  =
  \bigg\langle
  \bigg(
  \sum_t
  \Big\{
  \frac{\big(dt f(x_t)+g(x_t)\eta_tdt\big)^4}{dt}
  -
 3\,\big(2D g(x_t)^2\big)^2 \, dt
  \Big\}
  \bigg)^2
  \bigg\rangle
  \nonumber
  \\
&  \stackrel{\phantom{(\ref{eq:langevin_alpha})}}
  =
  \bigg\langle
  \bigg(
  \sum_t
  \Big\{
  \frac{\big(g(x_t)\eta_tdt\big)^4}{dt}
  -
 3\,\big(2D g(x_t)^2\big)^2 \, dt
  \Big\}
  \bigg)^2
  \bigg\rangle
  +O(dt)
  \label{eq:Oks4_f3a}
  \\
&  
  \stackrel{\phantom{(\ref{eq:langevin_alpha})}}
  =
  \bigg\langle
  \Big(
  \sum_t
   \big[(B_{t+dt}-B_t\big)^4 - 3 dt^2\big] \big(2D g(x_t)^2\big)^2dt^{-1}
  \Big)^2
  \bigg\rangle
  +O(dt)
  \label{eq:Oks4_1}
  \\
&
  \stackrel{\phantom{(\ref{eq:langevin_alpha})}}
  =\ldots
  \ \textnormal{as in~\eqref{eq:indepdB2}, using~(\ref{eq:indepBt}) and the average~(\ref{eq:vmB4})}
\nonumber  \\
&  
  \stackrel{\phantom{(\ref{eq:langevin_alpha})}}
  =
  \sum_t
  \underbrace
  {
  \Big\langle
  \big[(B_{t+dt}-B_t\big)^4 - 3 dt^2\big]^2
  \Big\rangle
  }
  _
  {
  =105\,dt^4-2\times 3\times 3\, dt^4+9\,dt^4
  }
  \Big\langle
  \big(2D g(x_t)^2\big)^4dt^{-2}
  \Big\rangle
  +O(dt)
  \label{eq:Oks4_2}
  \\
&  
  \stackrel{\phantom{(\ref{eq:langevin_alpha})}}
  =
  dt
  \sum_t
  96\, dt
  \Big\langle
  \big(2D g(x_t)^2\big)^4
  \Big\rangle
  +O(dt)
  \nonumber
  \\
 &  \stackrel{\phantom{(\ref{eq:vmB2})}}
  =
   O(dt)
   \label{eq:finaldt_O2_4}
\end{align}
\end{small}%
which goes to zero as $dt\to 0$, hence finishing the proof of~(\ref{eq:Ito4}).
The derivations of~(\ref{eq:Ito3}) and~(\ref{eq:Ito6}) follow in the same way.

\medskip
Note that in passing from~(\ref{eq:Oks4_1}) to~(\ref{eq:Oks4_2}) 
we have used (i) the same independence as in~(\ref{eq:factorBrown}) and (ii) the fact that in the double sum term
\begin{align}
  \sum_{t\neq t'}
  \
  \bigg\langle
  &
   \big[(B_{t+dt}-B_t\big)^4 - 3 dt^2\big] \big(2D g(x_t)^2\big)^2dt^{-1}
  \bigg\rangle  
    \nonumber
  \\
  &
    \times
  \bigg\langle
   \big[(B_{t'+dt}-B_{t'}\big)^4 - 3 dt^2\big] \big(2D g(x_{t'})^2\big)^2dt^{-1}
  \bigg\rangle  
  \label{eq:doublesumdetails_Oks4}
\;,
\end{align}
which is similar to~(\ref{eq:indepdB2}), one again has the important cancellation
\begin{align}
  \label{eq:details_Oks4}
\fl\qquad
  \sum_t
  \bigg\langle
    \big[(B_{t+dt}&-B_t\big)^4 - 3 dt^2\big] \big(2D g(x_t)^2\big)^2dt^{-1}
  \bigg\rangle  
  \nonumber\\
  &                               
    \stackrel{\phantom{(\ref{eq:vmB4})}}
    =
    \bigg\langle
  \sum_t
    \big[(B_{t+dt}-B_t\big)^4 - 3 dt^2\big]
    \bigg\rangle
    \bigg\langle
\big(2D g(x_t)^2\big)^2dt^{-1}
    \bigg\rangle
\\
&  \stackrel{\text{(\ref{eq:vmB4})}}
  =
  0
  \label{eq:zero_average4}
\;.
\end{align}
In particular, the factor $3$ in the substitution rule $\Delta x^4 dt^{-1} = 3\: \big(2D g(x)^2\big)^2 \,dt$ one wants to show is essential, because it allows one to factorise by $2D g(x)^2$ between~(\ref{eq:Oks4_f3a}) and~(\ref{eq:Oks4_1}), and to obtain \emph{in fine} the cancellation in~\eqref{eq:zero_average4} which makes that~\eqref{eq:finaldt_O2_4} is of order $dt$.
The factor $3$ in $\Delta x^4 dt^{-1} = 3\: \big(2D g(x)^2\big)^2 \,dt$  is thus exactly the same as the one, obtained \emph{e.g.}~à la Wick in~(\ref{eq:vmB4}).

\renewcommand{\thesection}{C}
\section{An inconsistency arising when applying the standard chain rule inside the dynamical action}
\label{sec:inconsistency_action_chain-rule}
%
%
%

We detail in this appendix how an invalid use of the standard stochastic chain rule~(\ref{eq:chainrule_example}) can lead to an inconsistency when changing variables in the dynamical action corresponding to the Langevin equation~(\ref{eq:langevin_alpha}).
This appendix is the translation to the Onsager--Machlup action of the 
App.~E of~\cite{aron_dynamical_2014_arxiv1} (version v1 of the arXiv preprint)
where the same inconsistency was observed in the Martin--Siggia--Rose--Janssen--De\,Dominicis formulation of the dynamical action.

We compare the direct path (downwards, on the left) of the commutative diagram represented on Fig.~\ref{fig:commutative-diag_non-lin-variable-change}, and the indirect path where one first (top arrow) changes variables from $x(t)$ to $U(t)=u(x(t))$ in the Langevin equation, then (right arrow downwards) constructs 
the action, and finally (down arrow leftwards) tries to come back to the Onsager--Machlup action by applying the standard stochastic chain rule~(\ref{eq:chainrule_example}).
On the way, one should not forget to handle correctly the change of variables in the normalisation prefactor of the action.

The direct path leads to the expression~(\ref{eq:OMactionalpha}) of the dynamical action, together with its associated normalisation prefactor~(\ref{eq:Jacobialpha}).
The indirect path starts by obtaining the Langevin equation~(\ref{eq:langevin_alpha_uU}) on $U(t)=u(x(t))$ and continues by writing the corresponding the Onsager--Machlup weight~(\ref{eq:OM_U_alpha}). The last step consists in attempting to come back to the Onsager--Machlup weight for the process $x(t)$ by a change of variables in the action and in the Jacobian.

\subsection{The normalisation prefactor}

One can focus on the first time step $0 \curvearrowright dt$ without loss of generality.
The change of variables in the normalisation prefactor involves two stages: (\emph{i}) taking into account the factor $u'(x_{dt})$ of~(\ref{eq:change_var}) that comes from the change of measure
and (\emph{ii}) actually passing from the variable $\bar U_0$ [given by Eq.~(\ref{eq:Ubar0})] to the variable $\bar x_0$ in the prefactor ${\mathcal N}/{|G(\bar U_0)|}$ of the  Onsager--Machlup weight~(\ref{eq:OM_U_alpha}).
Denoting by $J_{U}(\bar x_0)$ the elementary normalisation prefactor coming from this procedure, one has
\begin{align}
J_{U}(\bar x_0)
&
=
\frac{\mathcal N}{|G(\bar U_0)|}\,
u'(x_{dt})
\\
&
=
\frac{\mathcal N}{|g(\bar x_0)|}
\,
\left(
u'(x_{dt})
\,
\frac{|g(\bar x_0)|}{|G(\bar U_0)|}
\right)
\; ,
\label{eq:decompJx0bar}
\end{align}
where on the second line we have put in prefactor the expected contribution ${\mathcal N}/{|g(\bar x_0)|}$ of the first time step in the total path-integral normalisation prefactor~(\ref{eq:Jacobialpha}) on $x(t)$.
The other factor~${u'(x_{dt})\,|g(\bar x_0)|}/{|G(\bar U_0)|}$ gives a contribution that has to be incorporated into the exponential part of the infinitesimal propagator (\emph{i.e.}~into the action of the path integral in the $dt\to 0$ limit). 
To do so, one expresses $\bar U_0$ and $x_{dt}$ in terms of $\bar x_0$ and $\Delta x $ by means of Eqs.~(\ref{eq:Ubar0})-(\ref{eq:Xdt}) and one expands the result up to order $dt$, keeping in mind that $\Delta x=O(dt^{1/2})$.
One obtains from~\eqref{eq:decompJx0bar} that
\\
\begin{minipage}{1.0\linewidth}
\begin{align}
\fl
J_{U}(\bar x_0)
=
&
\!\!
\frac{\mathcal N}{|g(\bar x_0)|}
\;
\bigg[
1-\frac{(-1+\alpha ) u''(\bar{x}_0)}{u'(\bar{x}_0)} \Delta x 
+
\label{eq:decompJx0bar_expanded}
\\
\!
\fl
&
\qquad\quad\,
(-1+\alpha )
\bigg(
\frac{ \alpha  g'(\bar{x}_0) u''(\bar{x}_0)}{2 g(\bar{x}_0) u'(\bar{x}_0)}
+\frac{ \alpha  u''(\bar{x}_0){}^2}{2 u'(\bar{x}_0){}^2}
+\frac{(-1+\alpha ) u^{(3)}(\bar{x}_0)}{2 u'(\bar{x}_0)}
\bigg)
\Delta x^2
\bigg]
\; . 
\nonumber
\end{align}%
\end{minipage}%
\\[3mm]
Then using the substitution rule~(\ref{eq:itopresc}) for $\Delta x^2$ and reexponentiating the result through~(\ref{eq:devexpWN}) one gets
\\
\begin{minipage}{1.0\linewidth}%
\begin{small}%
\begin{align}
\fl
J_{U}(\bar x_0)
=
\frac{\mathcal N}{|g(\bar x_0)|}
\;
\exp
\!
\bigg\{
&
-\frac{(-1+\alpha )  u''(\bar{x}_0)}{u'(\bar{x}_0)} \Delta x
\nonumber
\\
\fl
&
+
\bigg[
\frac{D (-1+\alpha ) g(\bar{x}_0){}^2 u''(\bar{x}_0){}^2}{u'(\bar{x}_0){}^2}
+
\label{eq:decompJx0bar_expanded_exponent}
\\
\fl
&
\quad\;\;\,
\frac{D (-1+\alpha ) g(\bar{x}_0) [\alpha  g'(\bar{x}_0) u''(\bar{x}_0)+(-1+\alpha ) g(\bar{x}_0) u^{(3)}(\bar{x}_0)]}{u'(\bar{x}_0)}
\bigg] 
dt
\bigg\}
\; . 
\nonumber
\end{align}%
\end{small}%
\end{minipage}%
\\[3mm]
Taking finally the continuous-time limit, we see that the change of variables from $U(t)$ to $x(t)$ brings a contribution $\Delta S_U[x(t)]$ to the action equal to
\\
\begin{minipage}{1.0\linewidth}
\begin{align}
\fl
\Delta S_U[x(t)]
\stackrel\alpha
=
\int_0^{\tf}
\!\!
dt\:
\bigg\{
&
\frac{(-1+\alpha )  u''(x)}{u'(x)} \:\dt x
-
\frac{D (-1+\alpha ) g(x){}^2 u''(x){}^2}{u'(x){}^2}
\nonumber
\\
\fl
&
-
\frac{D (-1+\alpha ) g(x) \big[\alpha  g'(x) u''(x)+(-1+\alpha ) g(x) u^{(3)}(x)\big]}{u'(x)}
\bigg\}
\;.
\label{eq:contrib_jacob_action}
\end{align}%
\end{minipage}%
\\[3mm]
We note that it vanishes for a linear transformation such that $u''=0$, or for $\alpha=1$ and any function $u$.
This last case is understood from~(\ref{eq:decompJx0bar}), where for $\alpha=1$ one has $\bar x_0=x_{dt}$ and $\bar U_0=U_{dt}=u(x_{dt})$ which implies, using~(\ref{eq:langevin_alpha_uU}) for the expression of $G$, that $G(\bar U_0)=u'(x_{dt}) g\big(u(x_{dt})\big)$ from which the factor in parenthesis in~(\ref{eq:decompJx0bar}) is equal to~1.

\subsection{The change of variables in the action}
\label{sec:action_wrong-chain-rule}

We can now combine the contribution $\Delta S_U[x(t)]$ obtained in the previous subsection and the change of variables from $U(t)=u(x(t))$ to $x(t)$ in the action $S[U(t)]$ of the process~$U(t)$. The expression of~$S[U(t)]$ is read from~(\ref{eq:OM_U_alpha}).
The correct procedure to follow is discussed in subsec.~\ref{sec:non-line-transf} following a discrete-time approach.
One can also apply the continuous-time modified chain-rule discussed in subsec.~\ref{sec:modified-chain-rule}.
Both approaches yield back the correct action~(\ref{eq:OMactionalpha}) for the process~$x(t)$.

If one  improperly applies 
the chain rule~(\ref{eq:chainrule_example}) to determine $\dt U=\dt\big[u(x(t))\big]$ in~(\ref{eq:OM_U_alpha}), one finds  a result  for the action $S_U\big[x(t)]$, 
in which there are supplementary terms compared do the correct result $S[x(t)]$ given by~(\ref{eq:OMactionalpha}); that is
\begin{align}
\fl
  S_U\big[x(t)]
&
  \stackrel\alpha
  =
  \Delta S_U[x(t)]
  +
  S\big[U(t)\big]\Big|_{U(t)=u(x(t))}
\\
\fl
&
  \stackrel\alpha
  =
S[x(t)]
+
\int_0^{\tf}\!\! dt\:
\bigg\{
\frac{(-1+2 \alpha )  u''(x)}{u'(x)} \:\dt x
-
\frac{3 D (-1+\alpha ) \alpha  g(x) g'(x) u''(x)}{u'(x)}
\nonumber
\\
\fl
&
\hspace*{31mm}
+
\frac{D (1+(-1+\alpha ) \alpha ) g(x)^2 u''(x)^2}{u'(x)^2}
\nonumber
\\
\fl
&
\hspace*{31mm}
+
\frac{D (-1-3 (-1+\alpha ) \alpha ) g(x)^2 u^{(3)}(x)}{u'(x)}
\bigg\}
\; . 
\label{eq:DeltaSU}
\end{align}
The terms in the time integral should be absent if the procedure had been correct. (At the very least, the result should not depend on the function $u$.)
These terms are equivalent to the terms in Eq.~(E.18) of the App.~E in~\cite{aron_dynamical_2014_arxiv1}.
Their presence is due to the fact that, when using the chain rule~(\ref{eq:chainrule_example}) as we did, one discards terms proportional to $\Delta x^3 dt^{-1}$, $\Delta x^4 dt^{-1}$ and $\Delta x^6 dt^{-2}$ that are present in Eq.~(\ref{eq:resdev}) when following the correct procedure.
The supplementary terms vanish when a linear change of variables is applied, \emph{i.e.}~when $u''=0$.

\emph{Special cases{~--~}} 
One notes that these supplementary terms still remain present in the three following simplified cases:
\begin{enumerate}
\item Stratonovich discretisation ($\alpha=1/2$):
\begin{align}
\fl
  S_U\big[x(t)]
&
  \stackrel\st
  =
S[x(t)]
+
\int_0^{\tf}\!\! dt\:
\bigg\{
\frac{3 D g(x) g'(x) u''(x)}{4 u'(x)}
+
\frac{3 D g(x)^2 u''(x)^2}{4 u'(x)^2}
\nonumber
\\
\fl
&
\hspace*{60mm}
-\frac{D g(x)^2 u^{(3)}(x)}{4 u'(x)}
\bigg\}
\end{align}

\item Additive noise ($g(x)=g$ is a constant):
\begin{align}
\fl
  S_U\big[x(t)]
&
  \stackrel\alpha
  =
S[x(t)]
+
\int_0^{\tf}\!\! dt\:
\bigg\{
\frac{(-1+2 \alpha ) u''(x)}{u'(x)}\dt x
+
\frac{D g^2 (1+(-1+\alpha ) \alpha ) u''(x)^2}{u'(x)^2}
\nonumber
\\
\fl
&
\hspace*{31mm}
+
\frac{D g^2 (-1-3 (-1+\alpha ) \alpha ) u^{(3)}(x)}{u'(x)}
\bigg\}
\end{align}

\item Additive noise and Stratonovich discretisation:
\begin{align}
\fl
  S_U\big[x(t)]
&
  \stackrel\st
  =
S[x(t)]
+
\int_0^{\tf}\!\! dt\:
D g^2
\bigg\{
\frac{3 u''(x)^2}{4 u'(x)^2}-\frac{ u^{(3)}(x)}{4 u'(x)}
\bigg\}
\end{align}
This last case is surprising because, as often described, the additive-noise Stratonovich-discretised Langevin equation is the better behaved in terms of the rules of differential calculus.
But in spite of this fact, as we have shown, the standard chain rule of differential calculus \emph{cannot} be used inside the corresponding Onsager--Machlup action (although this rule is valid at the Langevin equation level).

\end{enumerate}

\renewcommand{\thesection}{D}
\section{An inconsistency arising when applying the Langevin rule for changing discretisation inside the dynamical action}
\label{sec:inconsistency_action_ch-discretis}
%
%
%

In this appendix, we study how the $\alpha$-discretised Langevin equation~(\ref{eq:langevin_alpha}) can be described by a path-integral probability written in a 
different  $\bar\alpha$-discretisation.
The direct procedure to follow is to change the discretization in the Langevin equation first [this yields~(\ref{eq:langevin_alpha-to-alphabar}) with a modified force $f_{\alpha\to\bar\alpha} (x)$ given by~(\ref{eq:langevin_alpha-to-alphabar_force})], and to write the corresponding trajectory weight. One reads its action from~(\ref{eq:OMactionalpha}) as
\begin{align}
\label{eq:OMactionalpha-alphabar}
\fl\quad
  S_{\alpha\bar\alpha}[x(t)]
&  \stackrel{\bar\alpha}=
  \int_0^{\tf}\! dt\:
  \bigg\{
  \frac 12 \frac{1}{2D} 
  \bigg[\frac
  {\dt x-f_{\alpha\to\bar\alpha} (x)+2\bar\alpha D \,g(x)g'(x)}
  {g(x)}\bigg]^2
  +
  \bar\alpha  f'_{\alpha\to\bar\alpha}(x)
\bigg\}
\:,
\end{align}
where the arguments of the functions $f$ and $g$ are taken in $\bar\alpha$-discretisation.
The associated normalisation prefactor reads, from~(\ref{eq:Jacobialpha})
\begin{align}
  \mathcal J[x(t)]
  \stackrel{\bar\alpha}=
  \prod_t 
\bigg\{
  \sqrt{\frac{dt^{-1}}{4\pi D}}
  \frac
  1
  {|g(\bar x_t)|}
\bigg\}
\; . 
\label{eq:Jacobialpha-alphabar}
\end{align}

The correct way of performing the change in discretisation in the original $\alpha$-discretized action~(\ref{eq:OMactionalpha}) [together with the prefactor~(\ref{eq:Jacobialpha})]  was described in subsec.~\ref{sec:change-discr-infin}, going through the infinitesimal propagator in discrete time and using the generalised substitution rules~(\ref{eq:Ito2b})-(\ref{eq:Ito6}). The computation is done in the special case $\bar\alpha=1/2$ but also works for any value of $\bar\alpha$.

The naive procedure discussed in subsec.~\ref{sec:discussion_changeofdiscretisation}\hyperref[{sec:discussion_changeofdiscretisation}]{.d} consists in applying the two following steps.
\begin{itemize}
\item For the action: start from the ($\alpha$-discretized) action~(\ref{eq:OMactionalpha}) and assume that one 
can replace the force $f$ by the effective force $f_{\alpha\to\bar\alpha}$ appearing in the Langevin equation~(\ref{eq:langevin_alpha-to-alphabar}).
This would yield an exponential contribution to the action of the form
\begin{align}
\label{eq:OMactionalpha-alphabar-wrong}
\fl\qquad
  \int_0^{\tf}\! dt\:
  \bigg\{
  \frac 12 \frac{1}{2D} 
  \bigg[\frac
  {\dt x-f_{\alpha\to\bar\alpha} (x)+2\alpha D \,g(x)g'(x)}
  {g(x)}\bigg]^2
  +
  \bar\alpha  f'_{\alpha\to\bar\alpha}(x)
\bigg\}
\;.
\end{align}

\item For the normalisation prefactor: change the discretisation from $\alpha$ to $\bar \alpha$  by going to discrete time (and considering the first time step) and writing, with explicit discretisation points, that
  \begin{align}
\fl\qquad
\frac
1
{\big|g(\bar x_0^{(\alpha)})\big|}
=
\frac
1
{\big|g(\bar x_0^{(\bar\alpha)})\big|}
\frac
{\big|g(\bar x_0^{(\bar\alpha)})\big|}
{\big|g(\bar x_0^{(\alpha)})\big|}
\;.
\end{align}
Then, using
 $
   \bar x^{(\alpha)}_0
   = 
   \bar x_0^{(\bar\alpha)}+(\alpha-\bar\alpha)\Delta x
 $
to expand the second fraction in powers of $\Delta x= x_{dt}-x_0$ up to order $\Delta x^2$, using the substitution rule and reexponentiating the result gives
\begin{align}
\fl\ \
\frac
1
{\big|g(\bar x_0^{(\alpha)})\big|}
=
\frac
1
{\big|g(\bar x_0^{(\bar\alpha)})\big|}
\ee^{
\frac{  (-\alpha +\bar{\alpha }) g'(\bar x_0^{(\bar\alpha)})}{g(\bar x_0^{(\bar\alpha)})}\Delta x+D ( \alpha -\bar{\alpha })^2 \big[g'(\bar x_0^{(\bar\alpha)})^2-g(\bar x_0^{(\bar\alpha)}) g''(\bar x_0^{(\bar\alpha)})\big] dt
}
\;.
\end{align}
Taking the continuous-time limit, this gives the following contribution to the action (with $x$ being $\bar\alpha$-discretised):
\begin{align}
\fl\qquad
\int_0^{\tf}dt\;
\bigg\{
(\alpha -\bar{\alpha })
\frac{   g'(x)}{g(x)}\dt x - D ( \alpha -\bar{\alpha })^2 \big[g'(x)^2-g(x) g''(x) \big]
\bigg\}
\; . 
\label{eq:contribprefactorchdiscr}
\end{align}
Finally, adding~(\ref{eq:OMactionalpha-alphabar-wrong}) and~(\ref{eq:contribprefactorchdiscr}) yields a 
candidate $\tilde S_{\alpha\bar\alpha}[x(t)]$ for the $\bar\alpha$-discretised action of the $\alpha$-discretised Langevin equation.
If this procedure had been correct, one would have had recovered the same action $S_{\alpha\bar\alpha}[x(t)]$ as in~(\ref{eq:OMactionalpha-alphabar}), but, by direct inspection, one finds that
\begin{small}
\begin{align}
\fl\ \
\tilde S_{\alpha\bar\alpha}[x(t)]
=
&
S_{\alpha\bar\alpha}[x(t)]
\nonumber
\\
\fl
&
+\int_0^{\tf}dt\;
\bigg\{
\frac{(\alpha -\bar{\alpha }) \big[(-f(x)+2 \dt x) g'(x)+g(x) (f'(x)+2 D \bar{\alpha } g'(x)^2)\big]}{g(x)}
\nonumber
\\
\fl\ \
&
\hspace*{50mm}
+
3 D g(x) (\alpha -\bar{\alpha })^2 g''(x)
\bigg\}
\; .
\end{align}
\end{small}
\end{itemize}


\section*{References}

\addcontentsline{toc}{section}{References}
\bibliographystyle{plain_url}

\bibliography{langevin-calculus_equation-vs-action}
%


\end{document}